\documentclass[reprint,aps,prl,twocolumn]{revtex4-2}

\usepackage{graphicx}
\usepackage{amsmath}
\usepackage{amssymb}
\usepackage{dslabmath}
\usepackage{xcolor}
\usepackage{pdfpages}
\usepackage{atbegshi}

\makeatletter
\AtBeginDocument{\let\LS@rot\@undefined}
\makeatother

\begin{document}

\title{Information Entropy is a General-Purpose Collective Variable for Enhanced Sampling}

\author{Xiangrui Li}
\affiliation{Department of Materials Science and Engineering, University of California, Los Angeles, CA, United States}

\author{Daniel Schwalbe-Koda}
\email{dskoda@ucla.edu}
\affiliation{Department of Materials Science and Engineering, University of California, Los Angeles, CA, United States}

\begin{abstract}
\noindent
Enhanced sampling methods typically require predefined collective variables (CVs) that presuppose knowledge of reaction coordinates, restricting the discovery of unanticipated transition mechanisms or intermediates.
Here, we show that a local measure of information entropy in atomistic systems is a general-purpose CV for rare event sampling across molecular and condensed-phase systems.
The method biases simulations toward entropy-changing configurations following a well-tempered metadynamics approach, thus balancing novelty and thermodynamic accessibility.
Blind exploration of potential energy surfaces enables unsupervised discovery of metastable basins and reaction pathways, including competing transition channels inaccessible to conventional order parameters.
We demonstrate the generality of the method across five systems spanning conformational sampling, homogeneous nucleation, glass formation, and solid-state phase transformations.
\end{abstract}

\maketitle

Molecular dynamics (MD) simulations are widely used to study the kinetics and thermodynamics of molecules and materials, yet many processes of interest involve rare transitions across high-dimensional, rugged free energy landscapes inaccessible to unbiased MD simulations \cite{jansen2002concept, laio2002escaping, mannan2024navigating, pietrucci2017strategies}.
Enhanced sampling methods overcome this limitation by biasing the systems out of free energy minima along collective variables (CVs) that reduce the high-dimensional space into a small number of reaction coordinates \cite{torrie1977nonphysical, laio2002escaping, barducci2008well}.
However, the effectiveness of these methods depends on the careful selection of CVs that separate the relevant states and resolve the reaction pathways connecting them, often via physical or intuitive ansatzes \cite{palazzesi2017conformational, piaggi2017entropy}.
For instance, within nucleation and phase transformations, CVs are typically constructed from order parameters \cite{steinhardt1983bond} that classify local environments against a target, or coordination or pair-distribution-based descriptors \cite{gobbo2018nucleation,piaggi2017entropy} that quantify local ordering.
While effective, such order parameters make assumptions on the structure of the final state and may fail for systems of higher complexity or when multiple polymorphs compete \cite{piaggi2017enhancing}.

Machine learning (ML)-based CVs \cite{dietrich2023machine, bonati2021deep} or committor functions \cite{trizio2025everything} can handle more complex transitions in a data-driven way, but still depend on labeled structural or dynamical data from the relevant transition.
Recent works have also explored neural networks or model uncertainty as CVs \cite{chen2018collective,devergne2026seeds}, especially in the context of dataset construction \cite{kulichenko2023uncertainty,tan2025enhanced}.
Nevertheless, the quality of the CVs is dependent on quantities that are challenging to control, such as the reliability of uncertainty quantification methods or generalization capacity of ML models.
Finally, CVs defined for one system are rarely transferable to another, as even small perturbations in the chemical space can make them ineffective \cite{fu2024collective}.
A general-purpose CV should therefore simultaneously maximize resolution of sampled reaction pathways, minimize dependence on prior knowledge of the potential energy surface (PES), avoid models or training data, and favor low-energy pathways.

\begin{figure}[htb!]
    \centering
    \includegraphics[width=0.5\textwidth]{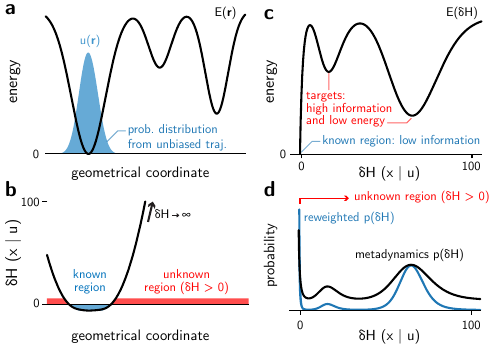}
    \caption{
    \textbf{a}, Example of a toy energy landscape projected onto a geometrical coordinate and the corresponding probability distribution $u(\textbf{r})$ of an unbiased trajectory trapped in the global energy minimum.
    \textbf{b}, $\dH$ as a function of geometrical coordinates given the distribution $u(\textbf{r})$.
    \textbf{c}, Energy landscape remapped from the geometrical to $\dH$ coordinate.
    \textbf{d}, Converged sampling probability from metadynamics simulation (black) and its reweighted Boltzmann probability distribution (blue) from remapped energy profile.
    }
    \label{fig:toy}
\end{figure}

Here, we propose that biasing simulations toward phase changes or unsampled configurations can be performed in a model-free approach using information entropy as a general CV for organic and inorganic systems alike.
Unlike structural entropy metrics that measure the degree of ordering of a configuration from pair correlations \cite{piaggi2017entropy, piaggi2017enhancing}, our approach quantifies the information content of each atomic environment irrespective of the type of structural order involved and without relying on training ML models.
Specifically, using the connection between the Shannon entropy $\Info = -\sum_{i} p_i \log_2 p_i$ \cite{shannon1948mathematical} and the thermodynamic entropy $S = -k_B \sum_{i} p_i \log p_i$, we bias simulations toward information entropy-changing configurations, thus creating distribution shifts on the probability space.
This approach therefore combines the enhanced sampling goal of operating directly on probability distributions \cite{invernizzi2020rethinking} with an information-theoretical view that estimates these distributions \cite{schwalbekoda2025information}.
To implement this method, we compute a differentiable, local information entropy change of an atomic environment represented as a high-dimensional vector $\Y \in \mathbb{R}^n$ with respect to a reference dataset $\Xset$, $\dH (\Y | \Xset) = -\log \sum_i K(\Y, \X_i)$, where $K$ is a kernel function \cite{schwalbekoda2025information}.
$\dH$ quantifies a measure of ``surprise'' of the environment $\Y$ given the reference $\Xset$, and thus can be used to distinguish between oversampled and new environments (Fig. \ref{fig:toy}b), with the absolute threshold $\dH \leq 0$ for $\Y \in \Xset$ and $\dH \rightarrow \infty$ for new samples.
Using this definition, we shift from sampling a distribution with respect to coordinates $\mathbf{r}$, $p \propto e^{-\beta E(\textbf{r})}$, where $\beta = 1 / k_B T$, and instead sample an information-theoretical probability landscape $p(\dH)$ inferred from a reference dataset $\Xset$ and energies $E(\dH)$ (Fig. \ref{fig:toy}c).
This avoids assumptions on target states, reaction pathways, geometrical coordinates, or ML models, and instead depends only on configurations $\Xset$ that can be trivially obtained from unbiased MD simulations or are computed on-the-fly from each MD snapshot.
Thus, this data-driven approach differs from structural entropy metrics that instead measure the degree of ordering of a structure \cite{piaggi2017entropy}.
In practice, $\Y$ and $\Xset$ can be built from any local atomic environment embedding method, as long as it can be implemented in a differentiable form to provide gradients for the simulations.
Here, we implemented a differentiable representation based on atom-centered symmetry functions \cite{behler2011atom} that provide both speed and reliability in representing environments, and a Gaussian kernel for $K$ in the definition of $\dH$, as implemented in the QUESTS approach \cite{schwalbekoda2025information}.
Enhanced sampling is thus implemented in an equivalent way to a metadynamics (MetaD) \cite{laio2002escaping} and well-tempered metadynamics (WT-MetaD) \cite{barducci2008well} with $\dH$ as the collective variable (Fig. \ref{fig:toy}d), balancing high-novelty and low-energy configurations \cite{invernizzi2022exploration}.
Figure \ref{fig:toy} illustrates this overall process by computing the distribution $u(\mathbf{r})$ of an unbiased simulation as the reference and the reaction coordinate is defined by $\dH \in [-\log N, +\infty)$, $N$ being the number of atoms of the system.
From the definition of $\dH$, the range for this CV is unvarying and interpretable, with $\dH \in [-\log N, 0]$ corresponding to well-sampled configurations, and $\dH > 0$ representing increasingly novel configurations given the reference distribution $u(\mathbf{r})$.

\begin{figure*}[htb!]
    \centering
    \includegraphics[width=1\textwidth]{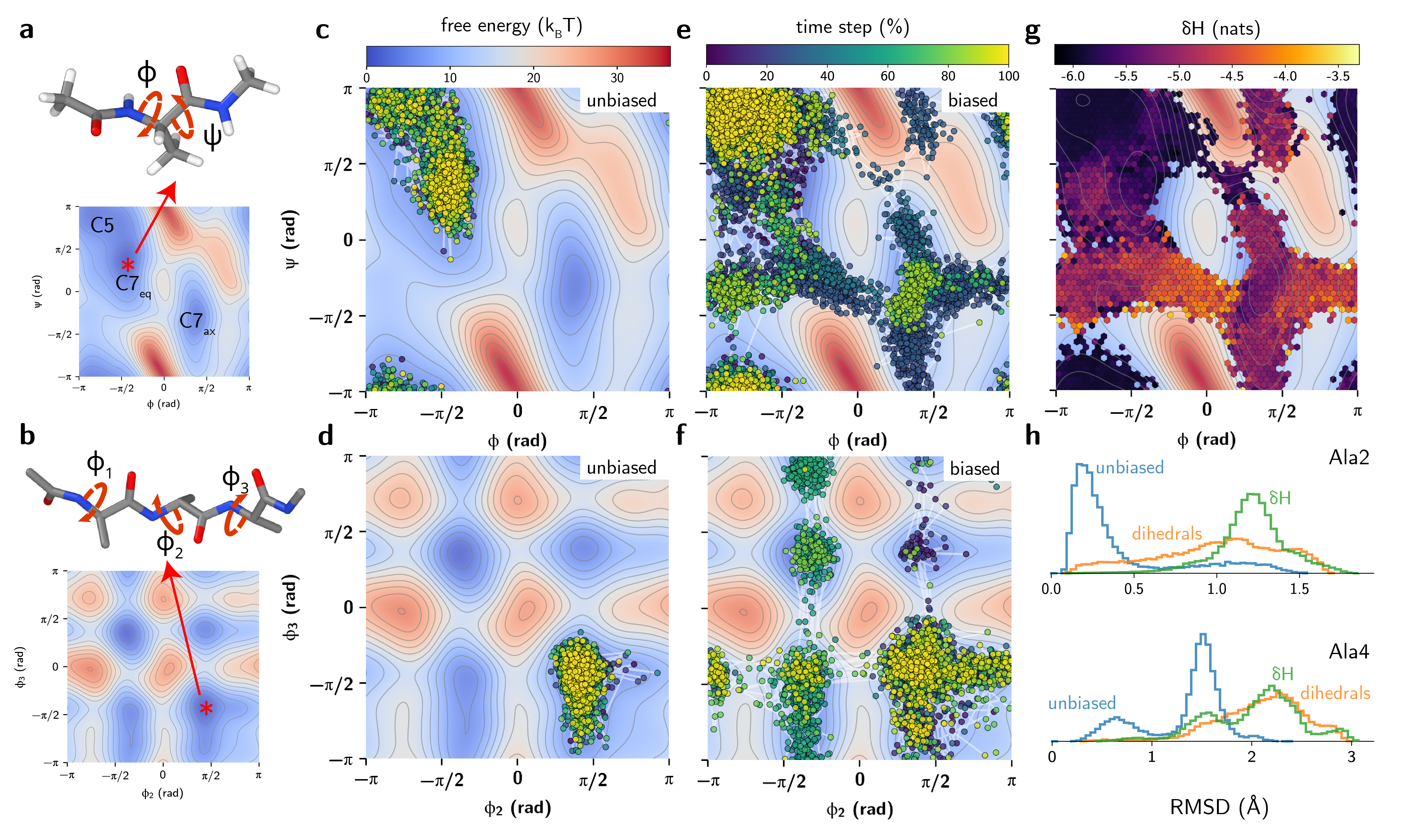}
    \caption{
        Structure and free energy surface of \textbf{a}, alanine dipeptide (Ala2) along the $(\phi, \psi)$ plane and \textbf{b}, alanine tetrapeptide (Ala4) along the $\phi_2$, $\phi_3$ plane.
        \textbf{c,d}, Unbiased simulation trajectories for Ala2 (\textbf{c}) and Ala4 (\textbf{d}).
        Initial structures are shown as red asterisks on the left, and colors on the right indicate the temporal order of sampled configurations.
        \textbf{e,f}, $\dH$-MetaD simulation trajectory using $\dH$ as CVs.
        \textbf{g}, Average $\dH$ distribution for the five atoms comprising $(\phi, \psi)$ in Ala2.
        \textbf{h}, Distribution of root mean square deviation (RMSD) of each frame in unbiased, dihedral-MetaD, and $\dH$-MetaD against the initial structure for Ala2 (top) and Ala4 (bottom).
    }
    \label{fig:organic}
\end{figure*}

We demonstrate the generality of our approach in organic and inorganic systems across conformational or phase transformations.
First, we validated $\dH$-MetaD on the well-known alanine dipeptide (Ala2) described by the rotation of backbone dihedral angles ($\phi, \psi$) known to be accurate CVs \cite{bolhuis2000reaction}.
We performed WT-MetaD in the $(\phi,\psi)$ space at $T=300$~K to obtain its free energy surface (FES) in vacuum (Fig.~\ref{fig:organic}a) with the Sage force field (v 2.0.0) in OpenFF \cite{boothroyd2023development, mobley2018escaping}, MD simulations using OpenMM \cite{eastman2023openmm} and PySAGES \cite{zubieta2024pysages}.
At 300~K, the unbiased trajectory remained trapped within the C5/C7eq basins and does not visit the C7ax state (Fig.~\ref{fig:organic}a,c).
Next, we randomly sampled 100 frames from an unbiased trajectory (300~K) as the reference dataset $\Xset$ and initialize the $\dH$-driven, WT-MetaD simulation in the C7eq basin (red asterisk in Fig. \ref{fig:organic}a).
As shown in Fig.~\ref{fig:organic}e, the biased simulation overcame the relevant energy barriers and samples the C7ax basin.
Importantly, the $\dH$-MetaD automatically discovered multiple energetically favorable pathways that connect metastable basins with no explicit guidance applied to dihedral rotation, demonstrating the effect of information-guided simulations.
To further elucidate this effect, Fig.~\ref{fig:organic}g reports the average per-atom $\dH$ computed for the five atoms defining the two dihedral angles.
Using the unbiased ensemble as the reference (Fig. \ref{fig:organic}c), configurations within the reference basins yield small $\dH$, whereas saddle regions and under-sampled basins exhibit relatively high novelty (larger $\dH$).
The estimated sampling probability distribution does not uniformly sample the space defined by $(\phi, \psi)$, but obtains the right pathways nevertheless.

The same workflow was demonstrated to alanine tetrapeptide (Ala4), which exhibits three backbone dihedrals ($\phi_1$, $\phi_2$, $\phi_3$) as natural variables \cite{hovan2018defining, tsai2021sgoop, devergne2026seeds}.
Its free energy surface in the $(\phi_2,\phi_3)$-projected plane (Fig.~\ref{fig:organic}b) showcases a more complex, multi-state system compared to Ala2 and poses higher sampling challenges without explicit definition of the torsion variables.
While an unbiased simulation at 300~K remains trapped in the initial basin (Fig.~\ref{fig:organic}d), the $\dH$-biased MetaD simulation samples all metastable states along the associated transition pathways (Fig.~\ref{fig:organic}f).
Sampled states in the three-dimensional torsional space (i.e., including $\phi_1$ as projection axis) show that $\dH$, despite being a one-dimensional CV, explores more states compared to two dihedrals and achieves comparable exploration coverage compared to the complete, three dihedrals CV set (Fig. S1).
To demonstrate how spaces are sampled differently with $\dH$ and geometrical CVs, Fig. \ref{fig:organic}h compares distributions of the root mean square deviation (RMSD) against the reference state selected from unbiased trajectory for Ala2 and Ala4 across sampling methods.
For both peptides, the $\dH$-driven method reaches a similar RMSD range as $(\phi, \psi)$-MetaD and very few samples at low RMSD, indicating its effectiveness at lowering the probability of reference environments compared to novel ones.
Notably, the RMSD histograms of $\dH$-MetaD in Fig. \ref{fig:organic}h exhibit peaks, equivalent to thermal fluctuations.
On the other hand, the simplified energy surface converges fast for dihedral-MetaD, but also samples high-energy pathways.
When the definitions of CV are incomplete with respect to the dimensionality of the sampled space, dihedral-MetaD does not explore other independent axes even where more energetic favorable pathways  exist (Fig. S1).

\begin{figure*}[htb!]
    \centering
    \includegraphics[width=1\textwidth]{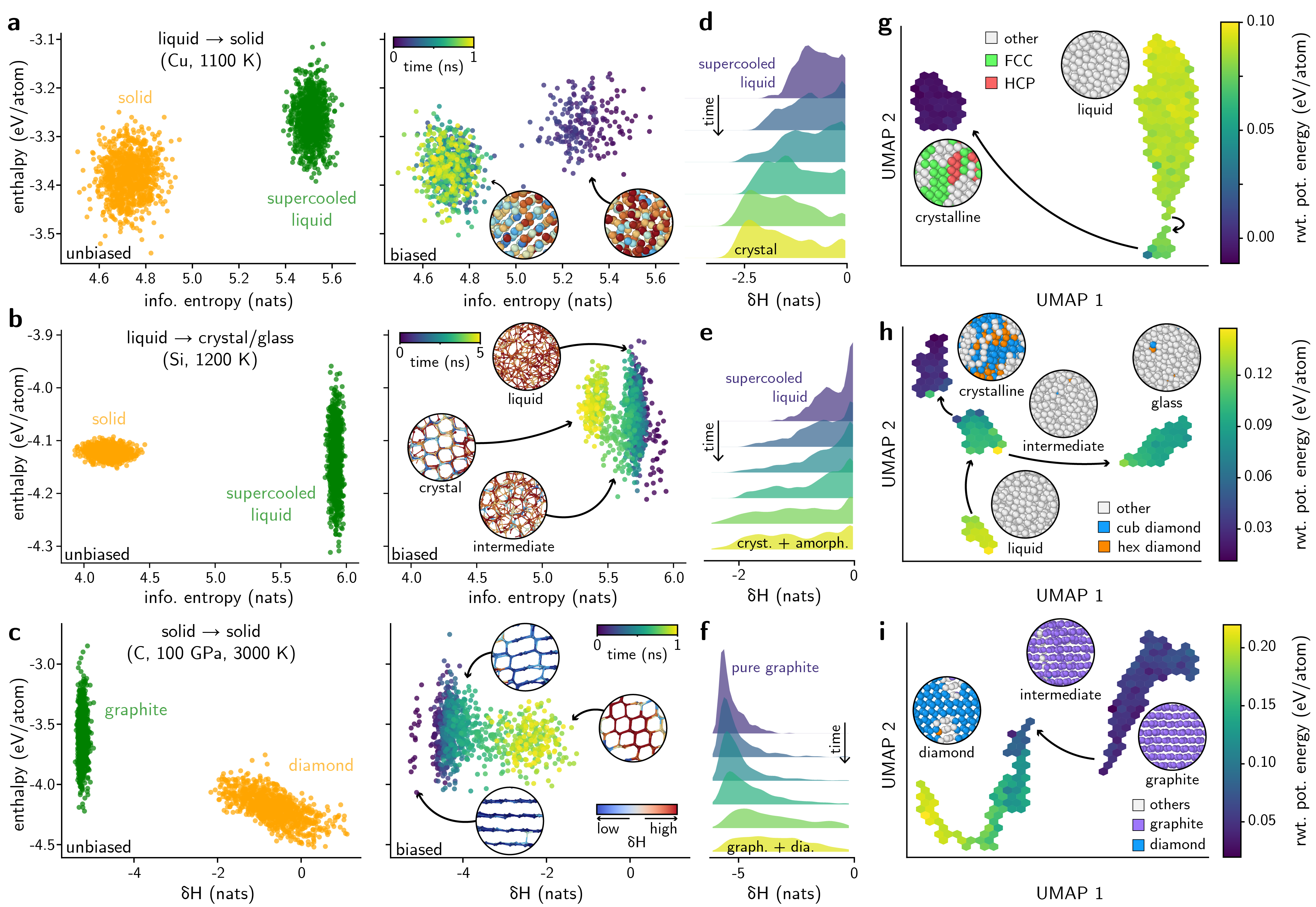}
    \caption{
        \textbf{a-c},
        \textbf{Left panel:} Enthalpy vs. information entropy diagram of two unbiased trajectories with different initial configurations for copper, silicon and carbon.
        \textbf{Right panel:} Enthalpy-information entropy diagram of biased trajectory.
        \textbf{d-f}, Time evolution of $\dH$ distribution shift over biased simulation trajectory.
        \textbf{g-i}, Boltzmann-reweighted potential energy profile on UMAP reduced dimension of Wasserstein-1 distance matrix of $\{\dH\}$ across multiple trajectories. Here $\{\dH\} = \{ \dH(\Y_{ij} | \Xset) \}$ is the set of $\dH$ computed over all frames $i$ and trajectories $j$.
    }
    \label{fig:inorganic}
\end{figure*}

Beyond conformational changes in molecules, we show that $\dH$ can sample kinetic transitions in inorganic materials along archetypical nucleation and phase transformation processes: nucleation of copper, nucleation or glass transition in silicon, and the graphite-to-diamond transformation in carbon, with MD simulations performed using LAMMPS \cite{thompson2022lammps} and PySAGES \cite{zubieta2024pysages}.
Similarly to CVs, order parameters are computed from mathematical operations of coordinate defined to distinguish crystalline status of inorganic materials and needs to be equivariant over translation and rotation \cite{steinhardt1983bond, neha2022collective, giberti2015metadynamics}.
Multiple works have been performed to characterize local atomic environments with descriptors \cite{steinhardt1983bond, piaggi2017enhancing} or ML \cite{dietrich2023machine}, but driving transformations often requires manual assignment of states or system-specific representations.
On the other hand, the information theoretical $\dH$ can distinguish between inorganic states without predefined thresholds.
For solid-state transformations, we extended the $\dH(\Y|\Xset)$ approach to periodic systems without loss of generality.
For order-disorder or disorder-disorder transformations, reference datasets are no longer needed, as each frame is its own reference state $\Yset$.
If a system is fully crystalline, then $\Y_i \approx \Y_j$ for any pair $\Y_i,\Y_j \in \Yset$, giving $\dH(\Y|\Yset) \approx -\log N$, where $N$ is the number of atoms.
In contrast, when local environments differ substantially (e.g., liquid or amorphous states), $Y_i \neq Y_j$ and $\dH(\Y|\Yset)\approx 0$.
As a result, disordered-to-ordered transitions can be mapped onto a bounded reaction coordinate $\dH(\Y|\Yset) \in [-\log N,\,0]$ without the need for reference states $\Xset$.
The total information entropy of the system, $\Info(\Yset) = \log{N} + \sum_{i=1}^N \dH(\Y_i|\Yset)$, also allows interpretation of the phases, with disordered phases mapping to high information entropy.

Copper solidification is a useful example of kinetic pathways in homogeneous nucleation \cite{sadigh2021metastable} and thus can be used to demonstrate our method for sampling order-disorder transformations with well-established classical force fields \cite{mishin2001structural}.
Since face centered cubic (FCC) copper is the most stable solid phase at 1 atm, we ran two, 10~ns-long unbiased simulations at 1100 K and 1 atm, but one starting from FCC and another from a liquid.
Both phases remained stable over this timescale due to the nucleation barrier and finite size effects, as illustrated by the enthalpy--information entropy plot in Fig.~\ref{fig:inorganic}a.
In contrast, in a biased simulation starting from a supercooled liquid, the system nucleates into a solid with coexisting FCC and hexagonal close-packed (HCP) motifs (Fig. \ref{fig:inorganic}a, right panel).
Furthermore, the $\dH$ distribution exhibits a clear shift from the disordered regime to the ordered regime during the transition (Fig.~\ref{fig:inorganic}d) and provides evidence for slight ordering of the supercooled liquid given the reasonable number of environments with $\dH < 0$, whereas a stable liquid often exhibits environments with mostly $\dH \sim 0$ \cite{schwalbekoda2025information}.
Figure \ref{fig:inorganic}g demonstrates this reaction pathway and FES in a two-dimensional (2D) space created from the $\dH$ distributions, further avoiding the definition of a geometrical CV (see Fig. S2 for 1D reaction pathway projected onto $\dH$).
The 2D space was defined by computing the pairwise Wasserstein-1 distances between each $\dH$ distribution across all $M$ frames to form a distance matrix $D\in\mathbb{R}^{M\times M}$, then using a dimensionality reduction method (UMAP) \cite{mcinnes2018umap} to visualize clusters along the reaction pathway in the $\dH$ space and correlations with the Boltzmann-reweighted potential energy (see Fig. S3 for consistency with a linear projection).
The analysis reveals three clusters along the solidification trajectory: a high-energy liquid region, a low-energy crystalline region, and a transient intermediate state (bottom right) comprised of a disordered state with local fluctuations in ordering measured by $\dH$ (Fig. S4).
Sampling this intermediate typical of non-classical nucleation mechanisms further demonstrates the ability of $\dH$-MetaD to induce phase transitions along favorable energy pathways \cite{lutsko2006theoretical}.

To showcase $\dH$-MetaD sampling on a more complex system, we performed MD simulations of silicon, which exhibits more sensitive phase transformation mechanisms across temperature and density \cite{beaucage2005nucleation}, including a glass transition.
As a model system, we performed simulations under the NVT ensemble with Stillinger-Weber (SW) potential \cite{stillinger1985computer} for a system in out-of-equilibrium conditions.
At 1200~K, a solid cubic-diamond silicon remained stable, whereas the supercooled liquid showed high instability, with most unbiased trajectories leading to crystallization within the simulated time scales.
On the other hand, biased simulations can either reproduce the crystallization pathway, exhibiting a shift in information entropy (Fig.~\ref{fig:inorganic}b,e), or steer the simulation towards a glass transition (Fig.~\ref{fig:inorganic}h).
As shown on the right panel of Fig.~\ref{fig:inorganic}h, the biased simulation samples a two-step nucleation process, first from a supercooled liquid to an amorphous intermediate captured by the $\dH$ (Fig. S5), and then to either a crystalline phase or a glassy state (Fig. S6).
This branching behavior is resolved despite the single scalar nature of the CV.
While unbiased simulations at these conditions consistently show that crystallization is the dominant reaction pathway, $\dH$-MetaD also samples a competing glass transition that is overlooked in this trajectory.
Importantly, sampling this transition would not be possible with structural ordering metrics that cannot distinguish between two disordered phases and are designed for order-disorder transformations.
This demonstrates that information entropy-driven sampling can resolve competing reaction channels without prior specification of the target phase.

Finally, to demonstrate that $\dH$-MetaD can sample solid-state phase transformations, we have tested our system on the graphite-diamond transformation in carbon, for which multiple experimental and computational pathways have been reported \cite{luo2024atomistic}.
Whereas entropy-like order parameters \cite{piaggi2017entropy} do not provide resolution to distinguish between crystalline phases, our information-theoretical approach allows us to separate them by adopting a single state as reference, similar to what was performed for Ala2 and Ala4.
The phase space exploration, however, remains blind with respect to the final state.
Using a Tersoff potential \cite{tersoff1988empirical}, this transformation occurs with at least $\sim$100~GPa and $\sim$3000~K, with cubic diamond being energetically less favorable than graphite \cite{marchant2023exploring}.
Independent, unbiased simulations initialized from both states show that phase transformations are not sampled at these conditions (Fig.~\ref{fig:inorganic}c).
On the other hand, a $\dH$-biased simulation using graphite as the reference dataset successfully samples the crystallization towards diamond, showcasing the slow nucleation and growth processes (Fig.~\ref{fig:inorganic}c) and shift of $\dH$ distributions from the reference graphite to the higher information diamond phase (Fig. \ref{fig:inorganic}f).
Contrary to the cases of copper or silicon, the 2D transformation pathway of Fig.~\ref{fig:inorganic}i shows only two dominant clusters from the low-energy graphite state to high-energy defective diamond state, as transient states (e.g., buckled graphite or shifts in stacking sequences) are closer to a continuous transformation than metastable intermediates according to an analysis of $\dH$ distributions.

In summary, we show that information entropy of atomistic environments is a general-purpose CV for blind exploration of phase spaces, as demonstrated by five case studies spanning conformational changes in organic molecules, and amorphous-to-crystalline, amorphous-to-amorphous, and crystalline-to-crystalline transformations.
Bias potentials placed at local, instantaneous values of $\dH$ quantifying the ``surprise'' of an environment estimate sampling probability with respect to a reference distribution, steering simulations towards low-probability, low-energy directions.
In practice, this novelty-seeking behavior means that independent runs may reveal distinct intermediates and energetically favorable pathways, as observed in transitions from Ala2 to the crystallization of silicon, allowing statistics to be obtained.
Blind exploration of metastable states without any previous knowledge is an ultimate goal from a mechanistic perspective \cite{devergne2026seeds, zhang2026exploring}, and can help reconstructing true probability surfaces in enhanced sampling \cite{gobbo2018nucleation, invernizzi2020rethinking, noe2019boltzmann, trizio2025everything}.
Our method provides a consistent formalism to estimate sampling probabilities directly from atomic environments while being agnostic to descriptors and devoid of ML models \cite{schwalbekoda2025information}.
Nevertheless, there are known trade-offs between convergence and exploration in CV-based enhanced sampling \cite{invernizzi2022exploration} which balance coverage in high-dimensional spaces and convergence speed in low-dimensional CV spaces.
While $\dH$ calculations can be adapted throughout the simulation by evolving the reference dataset $\Xset$, it may be unable to outperform well-defined reaction coordinates as CVs in terms of sampling speed.
One way to solve this problem is to include configurations sampled from a first, biased trajectory into the reference dataset that drives a second simulation, thus mimicking a two-step transformation with kinetic traps (Fig. S7).
This approach provides a starting point for simulations from which new CVs are selected, favoring non-directional exploration and enabling the discovery of previously unknown metastable states in the FES.
Given its generality, we anticipate this approach can advance blind sampling of rare events in problems ranging from enhanced sampling, kinetic Monte Carlo, molecular mechanics, and more.

\section*{Code and Data Availability}

The code used in this study will be publicly available upon publication.
The QUESTS package \cite{schwalbekoda2025information} is available at \url{https://github.com/dskoda/quests}.
OpenFF toolkit \cite{mobley2018escaping} was used to model the organic systems and it is available at \url{https://github.com/openforcefield/openff-toolkit}.
PySAGES \cite{zubieta2024pysages} is available at \url{https://github.com/SSAGESLabs/PySAGES}.

\section*{Acknowledgment}

This work was supported by the U.S. Department of Energy (DOE), Office of Science, Office of Basic Energy Sciences under Award Number DE-SC0025642.
This research used resources of the Argonne Leadership Computing Facility, which is a U.S. Department of Energy Office of Science User Facility operated under contract DE-AC02-06CH11357.

\section*{Conflicts of Interest}

The authors have no conflicts to disclose.

%apsrev4-2.bst 2019-01-14 (MD) hand-edited version of apsrev4-1.bst
%Control: key (0)
%Control: author (72) initials jnrlst
%Control: editor formatted (1) identically to author
%Control: production of article title (-1) disabled
%Control: page (0) single
%Control: year (1) truncated
%Control: production of eprint (0) enabled
%
% \bibliographystyle{apsrev4-2}
% \bibliography{refs-main}

\begin{thebibliography}{45}%
\makeatletter
\providecommand \@ifxundefined [1]{%
 \@ifx{#1\undefined}
}%
\providecommand \@ifnum [1]{%
 \ifnum #1\expandafter \@firstoftwo
 \else \expandafter \@secondoftwo
 \fi
}%
\providecommand \@ifx [1]{%
 \ifx #1\expandafter \@firstoftwo
 \else \expandafter \@secondoftwo
 \fi
}%
\providecommand \natexlab [1]{#1}%
\providecommand \enquote  [1]{``#1''}%
\providecommand \bibnamefont  [1]{#1}%
\providecommand \bibfnamefont [1]{#1}%
\providecommand \citenamefont [1]{#1}%
\providecommand \href@noop [0]{\@secondoftwo}%
\providecommand \href [0]{\begingroup \@sanitize@url \@href}%
\providecommand \@href[1]{\@@startlink{#1}\@@href}%
\providecommand \@@href[1]{\endgroup#1\@@endlink}%
\providecommand \@sanitize@url [0]{\catcode `\\12\catcode `\$12\catcode `\&12\catcode `\#12\catcode `\^12\catcode `\_12\catcode `\%12\relax}%
\providecommand \@@startlink[1]{}%
\providecommand \@@endlink[0]{}%
\providecommand \url  [0]{\begingroup\@sanitize@url \@url }%
\providecommand \@url [1]{\endgroup\@href {#1}{\urlprefix }}%
\providecommand \urlprefix  [0]{URL }%
\providecommand \Eprint [0]{\href }%
\providecommand \doibase [0]{https://doi.org/}%
\providecommand \selectlanguage [0]{\@gobble}%
\providecommand \bibinfo  [0]{\@secondoftwo}%
\providecommand \bibfield  [0]{\@secondoftwo}%
\providecommand \translation [1]{[#1]}%
\providecommand \BibitemOpen [0]{}%
\providecommand \bibitemStop [0]{}%
\providecommand \bibitemNoStop [0]{.\EOS\space}%
\providecommand \EOS [0]{\spacefactor3000\relax}%
\providecommand \BibitemShut  [1]{\csname bibitem#1\endcsname}%
\let\auto@bib@innerbib\@empty
%</preamble>
\bibitem [{\citenamefont {Jansen}(2002)}]{jansen2002concept}%
  \BibitemOpen
  \bibfield  {author} {\bibinfo {author} {\bibfnamefont {M.}~\bibnamefont {Jansen}},\ }\href@noop {} {\bibfield  {journal} {\bibinfo  {journal} {Angewandte Chemie International Edition}\ }\textbf {\bibinfo {volume} {41}},\ \bibinfo {pages} {3746} (\bibinfo {year} {2002})}\BibitemShut {NoStop}%
\bibitem [{\citenamefont {Laio}\ and\ \citenamefont {Parrinello}(2002)}]{laio2002escaping}%
  \BibitemOpen
  \bibfield  {author} {\bibinfo {author} {\bibfnamefont {A.}~\bibnamefont {Laio}}\ and\ \bibinfo {author} {\bibfnamefont {M.}~\bibnamefont {Parrinello}},\ }\href@noop {} {\bibfield  {journal} {\bibinfo  {journal} {Proceedings of the National Academy of Sciences}\ }\textbf {\bibinfo {volume} {99}},\ \bibinfo {pages} {12562} (\bibinfo {year} {2002})}\BibitemShut {NoStop}%
\bibitem [{\citenamefont {Mannan}\ \emph {et~al.}(2024)\citenamefont {Mannan}, \citenamefont {Bihani}, \citenamefont {Krishnan},\ and\ \citenamefont {Mauro}}]{mannan2024navigating}%
  \BibitemOpen
  \bibfield  {author} {\bibinfo {author} {\bibfnamefont {S.}~\bibnamefont {Mannan}}, \bibinfo {author} {\bibfnamefont {V.}~\bibnamefont {Bihani}}, \bibinfo {author} {\bibfnamefont {N.~A.}\ \bibnamefont {Krishnan}},\ and\ \bibinfo {author} {\bibfnamefont {J.~C.}\ \bibnamefont {Mauro}},\ }\href@noop {} {\bibfield  {journal} {\bibinfo  {journal} {Materials Genome Engineering Advances}\ }\textbf {\bibinfo {volume} {2}},\ \bibinfo {pages} {e25} (\bibinfo {year} {2024})}\BibitemShut {NoStop}%
\bibitem [{\citenamefont {Pietrucci}(2017)}]{pietrucci2017strategies}%
  \BibitemOpen
  \bibfield  {author} {\bibinfo {author} {\bibfnamefont {F.}~\bibnamefont {Pietrucci}},\ }\href@noop {} {\bibfield  {journal} {\bibinfo  {journal} {Reviews in Physics}\ }\textbf {\bibinfo {volume} {2}},\ \bibinfo {pages} {32} (\bibinfo {year} {2017})}\BibitemShut {NoStop}%
\bibitem [{\citenamefont {Torrie}\ and\ \citenamefont {Valleau}(1977)}]{torrie1977nonphysical}%
  \BibitemOpen
  \bibfield  {author} {\bibinfo {author} {\bibfnamefont {G.~M.}\ \bibnamefont {Torrie}}\ and\ \bibinfo {author} {\bibfnamefont {J.~P.}\ \bibnamefont {Valleau}},\ }\href@noop {} {\bibfield  {journal} {\bibinfo  {journal} {Journal of Computational Physics}\ }\textbf {\bibinfo {volume} {23}},\ \bibinfo {pages} {187} (\bibinfo {year} {1977})}\BibitemShut {NoStop}%
\bibitem [{\citenamefont {Barducci}\ \emph {et~al.}(2008)\citenamefont {Barducci}, \citenamefont {Bussi},\ and\ \citenamefont {Parrinello}}]{barducci2008well}%
  \BibitemOpen
  \bibfield  {author} {\bibinfo {author} {\bibfnamefont {A.}~\bibnamefont {Barducci}}, \bibinfo {author} {\bibfnamefont {G.}~\bibnamefont {Bussi}},\ and\ \bibinfo {author} {\bibfnamefont {M.}~\bibnamefont {Parrinello}},\ }\href@noop {} {\bibfield  {journal} {\bibinfo  {journal} {Physical Review Letters}\ }\textbf {\bibinfo {volume} {100}},\ \bibinfo {pages} {020603} (\bibinfo {year} {2008})}\BibitemShut {NoStop}%
\bibitem [{\citenamefont {Palazzesi}\ \emph {et~al.}(2017)\citenamefont {Palazzesi}, \citenamefont {Valsson},\ and\ \citenamefont {Parrinello}}]{palazzesi2017conformational}%
  \BibitemOpen
  \bibfield  {author} {\bibinfo {author} {\bibfnamefont {F.}~\bibnamefont {Palazzesi}}, \bibinfo {author} {\bibfnamefont {O.}~\bibnamefont {Valsson}},\ and\ \bibinfo {author} {\bibfnamefont {M.}~\bibnamefont {Parrinello}},\ }\href@noop {} {\bibfield  {journal} {\bibinfo  {journal} {The journal of Physical Chemistry Letters}\ }\textbf {\bibinfo {volume} {8}},\ \bibinfo {pages} {4752} (\bibinfo {year} {2017})}\BibitemShut {NoStop}%
\bibitem [{\citenamefont {Piaggi}\ and\ \citenamefont {Parrinello}(2017)}]{piaggi2017entropy}%
  \BibitemOpen
  \bibfield  {author} {\bibinfo {author} {\bibfnamefont {P.~M.}\ \bibnamefont {Piaggi}}\ and\ \bibinfo {author} {\bibfnamefont {M.}~\bibnamefont {Parrinello}},\ }\href@noop {} {\bibfield  {journal} {\bibinfo  {journal} {The Journal of Chemical Physics}\ }\textbf {\bibinfo {volume} {147}} (\bibinfo {year} {2017})}\BibitemShut {NoStop}%
\bibitem [{\citenamefont {Steinhardt}\ \emph {et~al.}(1983)\citenamefont {Steinhardt}, \citenamefont {Nelson},\ and\ \citenamefont {Ronchetti}}]{steinhardt1983bond}%
  \BibitemOpen
  \bibfield  {author} {\bibinfo {author} {\bibfnamefont {P.~J.}\ \bibnamefont {Steinhardt}}, \bibinfo {author} {\bibfnamefont {D.~R.}\ \bibnamefont {Nelson}},\ and\ \bibinfo {author} {\bibfnamefont {M.}~\bibnamefont {Ronchetti}},\ }\href@noop {} {\bibfield  {journal} {\bibinfo  {journal} {Physical Review B}\ }\textbf {\bibinfo {volume} {28}},\ \bibinfo {pages} {784} (\bibinfo {year} {1983})}\BibitemShut {NoStop}%
\bibitem [{\citenamefont {Gobbo}\ \emph {et~al.}(2018)\citenamefont {Gobbo}, \citenamefont {Bellucci}, \citenamefont {Tribello}, \citenamefont {Ciccotti},\ and\ \citenamefont {Trout}}]{gobbo2018nucleation}%
  \BibitemOpen
  \bibfield  {author} {\bibinfo {author} {\bibfnamefont {G.}~\bibnamefont {Gobbo}}, \bibinfo {author} {\bibfnamefont {M.~A.}\ \bibnamefont {Bellucci}}, \bibinfo {author} {\bibfnamefont {G.~A.}\ \bibnamefont {Tribello}}, \bibinfo {author} {\bibfnamefont {G.}~\bibnamefont {Ciccotti}},\ and\ \bibinfo {author} {\bibfnamefont {B.~L.}\ \bibnamefont {Trout}},\ }\href@noop {} {\bibfield  {journal} {\bibinfo  {journal} {Journal of Chemical Theory and Computation}\ }\textbf {\bibinfo {volume} {14}},\ \bibinfo {pages} {959} (\bibinfo {year} {2018})}\BibitemShut {NoStop}%
\bibitem [{\citenamefont {Piaggi}\ \emph {et~al.}(2017)\citenamefont {Piaggi}, \citenamefont {Valsson},\ and\ \citenamefont {Parrinello}}]{piaggi2017enhancing}%
  \BibitemOpen
  \bibfield  {author} {\bibinfo {author} {\bibfnamefont {P.~M.}\ \bibnamefont {Piaggi}}, \bibinfo {author} {\bibfnamefont {O.}~\bibnamefont {Valsson}},\ and\ \bibinfo {author} {\bibfnamefont {M.}~\bibnamefont {Parrinello}},\ }\href@noop {} {\bibfield  {journal} {\bibinfo  {journal} {Physical Review Letters}\ }\textbf {\bibinfo {volume} {119}},\ \bibinfo {pages} {015701} (\bibinfo {year} {2017})}\BibitemShut {NoStop}%
\bibitem [{\citenamefont {Dietrich}\ \emph {et~al.}(2023)\citenamefont {Dietrich}, \citenamefont {Advincula}, \citenamefont {Gobbo}, \citenamefont {Bellucci},\ and\ \citenamefont {Salvalaglio}}]{dietrich2023machine}%
  \BibitemOpen
  \bibfield  {author} {\bibinfo {author} {\bibfnamefont {F.~M.}\ \bibnamefont {Dietrich}}, \bibinfo {author} {\bibfnamefont {X.~R.}\ \bibnamefont {Advincula}}, \bibinfo {author} {\bibfnamefont {G.}~\bibnamefont {Gobbo}}, \bibinfo {author} {\bibfnamefont {M.~A.}\ \bibnamefont {Bellucci}},\ and\ \bibinfo {author} {\bibfnamefont {M.}~\bibnamefont {Salvalaglio}},\ }\href@noop {} {\bibfield  {journal} {\bibinfo  {journal} {Journal of Chemical Theory and Computation}\ }\textbf {\bibinfo {volume} {20}},\ \bibinfo {pages} {1600} (\bibinfo {year} {2023})}\BibitemShut {NoStop}%
\bibitem [{\citenamefont {Bonati}\ \emph {et~al.}(2021)\citenamefont {Bonati}, \citenamefont {Piccini},\ and\ \citenamefont {Parrinello}}]{bonati2021deep}%
  \BibitemOpen
  \bibfield  {author} {\bibinfo {author} {\bibfnamefont {L.}~\bibnamefont {Bonati}}, \bibinfo {author} {\bibfnamefont {G.}~\bibnamefont {Piccini}},\ and\ \bibinfo {author} {\bibfnamefont {M.}~\bibnamefont {Parrinello}},\ }\href@noop {} {\bibfield  {journal} {\bibinfo  {journal} {Proceedings of the National Academy of Sciences}\ }\textbf {\bibinfo {volume} {118}},\ \bibinfo {pages} {e2113533118} (\bibinfo {year} {2021})}\BibitemShut {NoStop}%
\bibitem [{\citenamefont {Trizio}\ \emph {et~al.}(2025)\citenamefont {Trizio}, \citenamefont {Kang},\ and\ \citenamefont {Parrinello}}]{trizio2025everything}%
  \BibitemOpen
  \bibfield  {author} {\bibinfo {author} {\bibfnamefont {E.}~\bibnamefont {Trizio}}, \bibinfo {author} {\bibfnamefont {P.}~\bibnamefont {Kang}},\ and\ \bibinfo {author} {\bibfnamefont {M.}~\bibnamefont {Parrinello}},\ }\href@noop {} {\bibfield  {journal} {\bibinfo  {journal} {Nature Computational Science}\ }\textbf {\bibinfo {volume} {5}},\ \bibinfo {pages} {582} (\bibinfo {year} {2025})}\BibitemShut {NoStop}%
\bibitem [{\citenamefont {Chen}\ \emph {et~al.}(2018)\citenamefont {Chen}, \citenamefont {Tan},\ and\ \citenamefont {Ferguson}}]{chen2018collective}%
  \BibitemOpen
  \bibfield  {author} {\bibinfo {author} {\bibfnamefont {W.}~\bibnamefont {Chen}}, \bibinfo {author} {\bibfnamefont {A.~R.}\ \bibnamefont {Tan}},\ and\ \bibinfo {author} {\bibfnamefont {A.~L.}\ \bibnamefont {Ferguson}},\ }\href@noop {} {\bibfield  {journal} {\bibinfo  {journal} {The Journal of Chemical Physics}\ }\textbf {\bibinfo {volume} {149}} (\bibinfo {year} {2018})}\BibitemShut {NoStop}%
\bibitem [{\citenamefont {Devergne}\ \emph {et~al.}(2026)\citenamefont {Devergne}, \citenamefont {Kostic}, \citenamefont {Pontil},\ and\ \citenamefont {Parrinello}}]{devergne2026seeds}%
  \BibitemOpen
  \bibfield  {author} {\bibinfo {author} {\bibfnamefont {T.}~\bibnamefont {Devergne}}, \bibinfo {author} {\bibfnamefont {V.}~\bibnamefont {Kostic}}, \bibinfo {author} {\bibfnamefont {M.}~\bibnamefont {Pontil}},\ and\ \bibinfo {author} {\bibfnamefont {M.}~\bibnamefont {Parrinello}},\ }\href@noop {} {\bibfield  {journal} {\bibinfo  {journal} {Proceedings of the National Academy of Sciences}\ }\textbf {\bibinfo {volume} {123}},\ \bibinfo {pages} {e2524602123} (\bibinfo {year} {2026})}\BibitemShut {NoStop}%
\bibitem [{\citenamefont {Kulichenko}\ \emph {et~al.}(2023)\citenamefont {Kulichenko}, \citenamefont {Barros}, \citenamefont {Lubbers}, \citenamefont {Li}, \citenamefont {Messerly}, \citenamefont {Tretiak}, \citenamefont {Smith},\ and\ \citenamefont {Nebgen}}]{kulichenko2023uncertainty}%
  \BibitemOpen
  \bibfield  {author} {\bibinfo {author} {\bibfnamefont {M.}~\bibnamefont {Kulichenko}}, \bibinfo {author} {\bibfnamefont {K.}~\bibnamefont {Barros}}, \bibinfo {author} {\bibfnamefont {N.}~\bibnamefont {Lubbers}}, \bibinfo {author} {\bibfnamefont {Y.~W.}\ \bibnamefont {Li}}, \bibinfo {author} {\bibfnamefont {R.}~\bibnamefont {Messerly}}, \bibinfo {author} {\bibfnamefont {S.}~\bibnamefont {Tretiak}}, \bibinfo {author} {\bibfnamefont {J.~S.}\ \bibnamefont {Smith}},\ and\ \bibinfo {author} {\bibfnamefont {B.}~\bibnamefont {Nebgen}},\ }\href@noop {} {\bibfield  {journal} {\bibinfo  {journal} {Nature Computational Science}\ }\textbf {\bibinfo {volume} {3}},\ \bibinfo {pages} {230} (\bibinfo {year} {2023})}\BibitemShut {NoStop}%
\bibitem [{\citenamefont {Tan}\ \emph {et~al.}(2025)\citenamefont {Tan}, \citenamefont {Dietschreit},\ and\ \citenamefont {G{\'o}mez-Bombarelli}}]{tan2025enhanced}%
  \BibitemOpen
  \bibfield  {author} {\bibinfo {author} {\bibfnamefont {A.~R.}\ \bibnamefont {Tan}}, \bibinfo {author} {\bibfnamefont {J.~C.}\ \bibnamefont {Dietschreit}},\ and\ \bibinfo {author} {\bibfnamefont {R.}~\bibnamefont {G{\'o}mez-Bombarelli}},\ }\href@noop {} {\bibfield  {journal} {\bibinfo  {journal} {The Journal of Chemical Physics}\ }\textbf {\bibinfo {volume} {162}} (\bibinfo {year} {2025})}\BibitemShut {NoStop}%
\bibitem [{\citenamefont {Fu}\ \emph {et~al.}(2024)\citenamefont {Fu}, \citenamefont {Bian}, \citenamefont {Shao},\ and\ \citenamefont {Cai}}]{fu2024collective}%
  \BibitemOpen
  \bibfield  {author} {\bibinfo {author} {\bibfnamefont {H.}~\bibnamefont {Fu}}, \bibinfo {author} {\bibfnamefont {H.}~\bibnamefont {Bian}}, \bibinfo {author} {\bibfnamefont {X.}~\bibnamefont {Shao}},\ and\ \bibinfo {author} {\bibfnamefont {W.}~\bibnamefont {Cai}},\ }\href@noop {} {\bibfield  {journal} {\bibinfo  {journal} {The Journal of Physical Chemistry Letters}\ }\textbf {\bibinfo {volume} {15}},\ \bibinfo {pages} {1774} (\bibinfo {year} {2024})}\BibitemShut {NoStop}%
\bibitem [{\citenamefont {Shannon}(1948)}]{shannon1948mathematical}%
  \BibitemOpen
  \bibfield  {author} {\bibinfo {author} {\bibfnamefont {C.~E.}\ \bibnamefont {Shannon}},\ }\href@noop {} {\bibfield  {journal} {\bibinfo  {journal} {The Bell System Technical Journal}\ }\textbf {\bibinfo {volume} {27}},\ \bibinfo {pages} {379} (\bibinfo {year} {1948})}\BibitemShut {NoStop}%
\bibitem [{\citenamefont {Invernizzi}\ and\ \citenamefont {Parrinello}(2020)}]{invernizzi2020rethinking}%
  \BibitemOpen
  \bibfield  {author} {\bibinfo {author} {\bibfnamefont {M.}~\bibnamefont {Invernizzi}}\ and\ \bibinfo {author} {\bibfnamefont {M.}~\bibnamefont {Parrinello}},\ }\href@noop {} {\bibfield  {journal} {\bibinfo  {journal} {The Journal of Physical Chemistry Letters}\ }\textbf {\bibinfo {volume} {11}},\ \bibinfo {pages} {2731} (\bibinfo {year} {2020})}\BibitemShut {NoStop}%
\bibitem [{\citenamefont {Schwalbe-Koda}\ \emph {et~al.}(2025)\citenamefont {Schwalbe-Koda}, \citenamefont {Hamel}, \citenamefont {Sadigh}, \citenamefont {Zhou},\ and\ \citenamefont {Lordi}}]{schwalbekoda2025information}%
  \BibitemOpen
  \bibfield  {author} {\bibinfo {author} {\bibfnamefont {D.}~\bibnamefont {Schwalbe-Koda}}, \bibinfo {author} {\bibfnamefont {S.}~\bibnamefont {Hamel}}, \bibinfo {author} {\bibfnamefont {B.}~\bibnamefont {Sadigh}}, \bibinfo {author} {\bibfnamefont {F.}~\bibnamefont {Zhou}},\ and\ \bibinfo {author} {\bibfnamefont {V.}~\bibnamefont {Lordi}},\ }\href {https://doi.org/10.1038/s41467-025-59232-0} {\bibfield  {journal} {\bibinfo  {journal} {Nature Communications}\ }\textbf {\bibinfo {volume} {16}},\ \bibinfo {pages} {4014} (\bibinfo {year} {2025})}\BibitemShut {NoStop}%
\bibitem [{\citenamefont {Behler}(2011)}]{behler2011atom}%
  \BibitemOpen
  \bibfield  {author} {\bibinfo {author} {\bibfnamefont {J.}~\bibnamefont {Behler}},\ }\href@noop {} {\bibfield  {journal} {\bibinfo  {journal} {The Journal of Chemical Physics}\ }\textbf {\bibinfo {volume} {134}} (\bibinfo {year} {2011})}\BibitemShut {NoStop}%
\bibitem [{\citenamefont {Invernizzi}\ and\ \citenamefont {Parrinello}(2022)}]{invernizzi2022exploration}%
  \BibitemOpen
  \bibfield  {author} {\bibinfo {author} {\bibfnamefont {M.}~\bibnamefont {Invernizzi}}\ and\ \bibinfo {author} {\bibfnamefont {M.}~\bibnamefont {Parrinello}},\ }\href@noop {} {\bibfield  {journal} {\bibinfo  {journal} {Journal of Chemical Theory and Computation}\ }\textbf {\bibinfo {volume} {18}},\ \bibinfo {pages} {3988} (\bibinfo {year} {2022})}\BibitemShut {NoStop}%
\bibitem [{\citenamefont {Bolhuis}\ \emph {et~al.}(2000)\citenamefont {Bolhuis}, \citenamefont {Dellago},\ and\ \citenamefont {Chandler}}]{bolhuis2000reaction}%
  \BibitemOpen
  \bibfield  {author} {\bibinfo {author} {\bibfnamefont {P.~G.}\ \bibnamefont {Bolhuis}}, \bibinfo {author} {\bibfnamefont {C.}~\bibnamefont {Dellago}},\ and\ \bibinfo {author} {\bibfnamefont {D.}~\bibnamefont {Chandler}},\ }\href@noop {} {\bibfield  {journal} {\bibinfo  {journal} {Proceedings of the National Academy of Sciences}\ }\textbf {\bibinfo {volume} {97}},\ \bibinfo {pages} {5877} (\bibinfo {year} {2000})}\BibitemShut {NoStop}%
\bibitem [{\citenamefont {Boothroyd}\ \emph {et~al.}(2023)\citenamefont {Boothroyd}, \citenamefont {Behara}, \citenamefont {Madin}, \citenamefont {Hahn}, \citenamefont {Jang}, \citenamefont {Gapsys}, \citenamefont {Wagner}, \citenamefont {Horton}, \citenamefont {Dotson}, \citenamefont {Thompson} \emph {et~al.}}]{boothroyd2023development}%
  \BibitemOpen
  \bibfield  {author} {\bibinfo {author} {\bibfnamefont {S.}~\bibnamefont {Boothroyd}}, \bibinfo {author} {\bibfnamefont {P.~K.}\ \bibnamefont {Behara}}, \bibinfo {author} {\bibfnamefont {O.~C.}\ \bibnamefont {Madin}}, \bibinfo {author} {\bibfnamefont {D.~F.}\ \bibnamefont {Hahn}}, \bibinfo {author} {\bibfnamefont {H.}~\bibnamefont {Jang}}, \bibinfo {author} {\bibfnamefont {V.}~\bibnamefont {Gapsys}}, \bibinfo {author} {\bibfnamefont {J.~R.}\ \bibnamefont {Wagner}}, \bibinfo {author} {\bibfnamefont {J.~T.}\ \bibnamefont {Horton}}, \bibinfo {author} {\bibfnamefont {D.~L.}\ \bibnamefont {Dotson}}, \bibinfo {author} {\bibfnamefont {M.~W.}\ \bibnamefont {Thompson}}, \emph {et~al.},\ }\href@noop {} {\bibfield  {journal} {\bibinfo  {journal} {Journal of Chemical Theory and Computation}\ }\textbf {\bibinfo {volume} {19}},\ \bibinfo {pages} {3251} (\bibinfo {year} {2023})}\BibitemShut {NoStop}%
\bibitem [{\citenamefont {Mobley}\ \emph {et~al.}(2018)\citenamefont {Mobley}, \citenamefont {Bannan}, \citenamefont {Rizzi}, \citenamefont {Bayly}, \citenamefont {Chodera}, \citenamefont {Lim}, \citenamefont {Lim}, \citenamefont {Beauchamp}, \citenamefont {Slochower}, \citenamefont {Shirts} \emph {et~al.}}]{mobley2018escaping}%
  \BibitemOpen
  \bibfield  {author} {\bibinfo {author} {\bibfnamefont {D.~L.}\ \bibnamefont {Mobley}}, \bibinfo {author} {\bibfnamefont {C.~C.}\ \bibnamefont {Bannan}}, \bibinfo {author} {\bibfnamefont {A.}~\bibnamefont {Rizzi}}, \bibinfo {author} {\bibfnamefont {C.~I.}\ \bibnamefont {Bayly}}, \bibinfo {author} {\bibfnamefont {J.~D.}\ \bibnamefont {Chodera}}, \bibinfo {author} {\bibfnamefont {V.~T.}\ \bibnamefont {Lim}}, \bibinfo {author} {\bibfnamefont {N.~M.}\ \bibnamefont {Lim}}, \bibinfo {author} {\bibfnamefont {K.~A.}\ \bibnamefont {Beauchamp}}, \bibinfo {author} {\bibfnamefont {D.~R.}\ \bibnamefont {Slochower}}, \bibinfo {author} {\bibfnamefont {M.~R.}\ \bibnamefont {Shirts}}, \emph {et~al.},\ }\href@noop {} {\bibfield  {journal} {\bibinfo  {journal} {Journal of Chemical Theory and Computation}\ }\textbf {\bibinfo {volume} {14}},\ \bibinfo {pages} {6076} (\bibinfo {year} {2018})}\BibitemShut {NoStop}%
\bibitem [{\citenamefont {Eastman}\ \emph {et~al.}(2023)\citenamefont {Eastman}, \citenamefont {Galvelis}, \citenamefont {Pel{\'a}ez}, \citenamefont {Abreu}, \citenamefont {Farr}, \citenamefont {Gallicchio}, \citenamefont {Gorenko}, \citenamefont {Henry}, \citenamefont {Hu}, \citenamefont {Huang} \emph {et~al.}}]{eastman2023openmm}%
  \BibitemOpen
  \bibfield  {author} {\bibinfo {author} {\bibfnamefont {P.}~\bibnamefont {Eastman}}, \bibinfo {author} {\bibfnamefont {R.}~\bibnamefont {Galvelis}}, \bibinfo {author} {\bibfnamefont {R.~P.}\ \bibnamefont {Pel{\'a}ez}}, \bibinfo {author} {\bibfnamefont {C.~R.}\ \bibnamefont {Abreu}}, \bibinfo {author} {\bibfnamefont {S.~E.}\ \bibnamefont {Farr}}, \bibinfo {author} {\bibfnamefont {E.}~\bibnamefont {Gallicchio}}, \bibinfo {author} {\bibfnamefont {A.}~\bibnamefont {Gorenko}}, \bibinfo {author} {\bibfnamefont {M.~M.}\ \bibnamefont {Henry}}, \bibinfo {author} {\bibfnamefont {F.}~\bibnamefont {Hu}}, \bibinfo {author} {\bibfnamefont {J.}~\bibnamefont {Huang}}, \emph {et~al.},\ }\href@noop {} {\bibfield  {journal} {\bibinfo  {journal} {The Journal of Physical Chemistry B}\ }\textbf {\bibinfo {volume} {128}},\ \bibinfo {pages} {109} (\bibinfo {year} {2023})}\BibitemShut {NoStop}%
\bibitem [{\citenamefont {Zubieta~Rico}\ \emph {et~al.}(2024)\citenamefont {Zubieta~Rico}, \citenamefont {Schneider}, \citenamefont {P{\'e}rez-Lemus}, \citenamefont {Alessandri}, \citenamefont {Dasetty}, \citenamefont {Nguyen}, \citenamefont {Men{\'e}ndez}, \citenamefont {Wu}, \citenamefont {Jin}, \citenamefont {Xu} \emph {et~al.}}]{zubieta2024pysages}%
  \BibitemOpen
  \bibfield  {author} {\bibinfo {author} {\bibfnamefont {P.~F.}\ \bibnamefont {Zubieta~Rico}}, \bibinfo {author} {\bibfnamefont {L.}~\bibnamefont {Schneider}}, \bibinfo {author} {\bibfnamefont {G.~R.}\ \bibnamefont {P{\'e}rez-Lemus}}, \bibinfo {author} {\bibfnamefont {R.}~\bibnamefont {Alessandri}}, \bibinfo {author} {\bibfnamefont {S.}~\bibnamefont {Dasetty}}, \bibinfo {author} {\bibfnamefont {T.~D.}\ \bibnamefont {Nguyen}}, \bibinfo {author} {\bibfnamefont {C.~A.}\ \bibnamefont {Men{\'e}ndez}}, \bibinfo {author} {\bibfnamefont {Y.}~\bibnamefont {Wu}}, \bibinfo {author} {\bibfnamefont {Y.}~\bibnamefont {Jin}}, \bibinfo {author} {\bibfnamefont {Y.}~\bibnamefont {Xu}}, \emph {et~al.},\ }\href@noop {} {\bibfield  {journal} {\bibinfo  {journal} {npj Computational Materials}\ }\textbf {\bibinfo {volume} {10}},\ \bibinfo {pages} {35} (\bibinfo {year} {2024})}\BibitemShut {NoStop}%
\bibitem [{\citenamefont {Hovan}\ \emph {et~al.}(2018)\citenamefont {Hovan}, \citenamefont {Comitani},\ and\ \citenamefont {Gervasio}}]{hovan2018defining}%
  \BibitemOpen
  \bibfield  {author} {\bibinfo {author} {\bibfnamefont {L.}~\bibnamefont {Hovan}}, \bibinfo {author} {\bibfnamefont {F.}~\bibnamefont {Comitani}},\ and\ \bibinfo {author} {\bibfnamefont {F.~L.}\ \bibnamefont {Gervasio}},\ }\href@noop {} {\bibfield  {journal} {\bibinfo  {journal} {Journal of Chemical Theory and Computation}\ }\textbf {\bibinfo {volume} {15}},\ \bibinfo {pages} {25} (\bibinfo {year} {2018})}\BibitemShut {NoStop}%
\bibitem [{\citenamefont {Tsai}\ \emph {et~al.}(2021)\citenamefont {Tsai}, \citenamefont {Smith},\ and\ \citenamefont {Tiwary}}]{tsai2021sgoop}%
  \BibitemOpen
  \bibfield  {author} {\bibinfo {author} {\bibfnamefont {S.-T.}\ \bibnamefont {Tsai}}, \bibinfo {author} {\bibfnamefont {Z.}~\bibnamefont {Smith}},\ and\ \bibinfo {author} {\bibfnamefont {P.}~\bibnamefont {Tiwary}},\ }\href@noop {} {\bibfield  {journal} {\bibinfo  {journal} {Journal of Chemical Theory and Computation}\ }\textbf {\bibinfo {volume} {17}},\ \bibinfo {pages} {6757} (\bibinfo {year} {2021})}\BibitemShut {NoStop}%
\bibitem [{\citenamefont {Thompson}\ \emph {et~al.}(2022)\citenamefont {Thompson}, \citenamefont {Aktulga}, \citenamefont {Berger}, \citenamefont {Bolintineanu}, \citenamefont {Brown}, \citenamefont {Crozier}, \citenamefont {In't~Veld}, \citenamefont {Kohlmeyer}, \citenamefont {Moore}, \citenamefont {Nguyen} \emph {et~al.}}]{thompson2022lammps}%
  \BibitemOpen
  \bibfield  {author} {\bibinfo {author} {\bibfnamefont {A.~P.}\ \bibnamefont {Thompson}}, \bibinfo {author} {\bibfnamefont {H.~M.}\ \bibnamefont {Aktulga}}, \bibinfo {author} {\bibfnamefont {R.}~\bibnamefont {Berger}}, \bibinfo {author} {\bibfnamefont {D.~S.}\ \bibnamefont {Bolintineanu}}, \bibinfo {author} {\bibfnamefont {W.~M.}\ \bibnamefont {Brown}}, \bibinfo {author} {\bibfnamefont {P.~S.}\ \bibnamefont {Crozier}}, \bibinfo {author} {\bibfnamefont {P.~J.}\ \bibnamefont {In't~Veld}}, \bibinfo {author} {\bibfnamefont {A.}~\bibnamefont {Kohlmeyer}}, \bibinfo {author} {\bibfnamefont {S.~G.}\ \bibnamefont {Moore}}, \bibinfo {author} {\bibfnamefont {T.~D.}\ \bibnamefont {Nguyen}}, \emph {et~al.},\ }\href@noop {} {\bibfield  {journal} {\bibinfo  {journal} {Computer Physics Communications}\ }\textbf {\bibinfo {volume} {271}},\ \bibinfo {pages} {108171} (\bibinfo {year} {2022})}\BibitemShut {NoStop}%
\bibitem [{\citenamefont {Neha}\ \emph {et~al.}(2022)\citenamefont {Neha}, \citenamefont {Tiwari}, \citenamefont {Mondal}, \citenamefont {Kumari},\ and\ \citenamefont {Karmakar}}]{neha2022collective}%
  \BibitemOpen
  \bibfield  {author} {\bibinfo {author} {\bibnamefont {Neha}}, \bibinfo {author} {\bibfnamefont {V.}~\bibnamefont {Tiwari}}, \bibinfo {author} {\bibfnamefont {S.}~\bibnamefont {Mondal}}, \bibinfo {author} {\bibfnamefont {N.}~\bibnamefont {Kumari}},\ and\ \bibinfo {author} {\bibfnamefont {T.}~\bibnamefont {Karmakar}},\ }\href@noop {} {\bibfield  {journal} {\bibinfo  {journal} {ACS Omega}\ }\textbf {\bibinfo {volume} {8}},\ \bibinfo {pages} {127} (\bibinfo {year} {2022})}\BibitemShut {NoStop}%
\bibitem [{\citenamefont {Giberti}\ \emph {et~al.}(2015)\citenamefont {Giberti}, \citenamefont {Salvalaglio},\ and\ \citenamefont {Parrinello}}]{giberti2015metadynamics}%
  \BibitemOpen
  \bibfield  {author} {\bibinfo {author} {\bibfnamefont {F.}~\bibnamefont {Giberti}}, \bibinfo {author} {\bibfnamefont {M.}~\bibnamefont {Salvalaglio}},\ and\ \bibinfo {author} {\bibfnamefont {M.}~\bibnamefont {Parrinello}},\ }\href@noop {} {\bibfield  {journal} {\bibinfo  {journal} {IUCrJ}\ }\textbf {\bibinfo {volume} {2}},\ \bibinfo {pages} {256} (\bibinfo {year} {2015})}\BibitemShut {NoStop}%
\bibitem [{\citenamefont {Sadigh}\ \emph {et~al.}(2021)\citenamefont {Sadigh}, \citenamefont {Zepeda-Ruiz},\ and\ \citenamefont {Belof}}]{sadigh2021metastable}%
  \BibitemOpen
  \bibfield  {author} {\bibinfo {author} {\bibfnamefont {B.}~\bibnamefont {Sadigh}}, \bibinfo {author} {\bibfnamefont {L.}~\bibnamefont {Zepeda-Ruiz}},\ and\ \bibinfo {author} {\bibfnamefont {J.~L.}\ \bibnamefont {Belof}},\ }\href@noop {} {\bibfield  {journal} {\bibinfo  {journal} {Proceedings of the National Academy of Sciences}\ }\textbf {\bibinfo {volume} {118}},\ \bibinfo {pages} {e2017809118} (\bibinfo {year} {2021})}\BibitemShut {NoStop}%
\bibitem [{\citenamefont {Mishin}\ \emph {et~al.}(2001)\citenamefont {Mishin}, \citenamefont {Mehl}, \citenamefont {Papaconstantopoulos}, \citenamefont {Voter},\ and\ \citenamefont {Kress}}]{mishin2001structural}%
  \BibitemOpen
  \bibfield  {author} {\bibinfo {author} {\bibfnamefont {Y.}~\bibnamefont {Mishin}}, \bibinfo {author} {\bibfnamefont {M.~J.}\ \bibnamefont {Mehl}}, \bibinfo {author} {\bibfnamefont {D.~A.}\ \bibnamefont {Papaconstantopoulos}}, \bibinfo {author} {\bibfnamefont {A.~F.}\ \bibnamefont {Voter}},\ and\ \bibinfo {author} {\bibfnamefont {J.~D.}\ \bibnamefont {Kress}},\ }\href@noop {} {\bibfield  {journal} {\bibinfo  {journal} {Physical Review B}\ }\textbf {\bibinfo {volume} {63}},\ \bibinfo {pages} {224106} (\bibinfo {year} {2001})}\BibitemShut {NoStop}%
\bibitem [{\citenamefont {McInnes}\ \emph {et~al.}(2018)\citenamefont {McInnes}, \citenamefont {Healy},\ and\ \citenamefont {Melville}}]{mcinnes2018umap}%
  \BibitemOpen
  \bibfield  {author} {\bibinfo {author} {\bibfnamefont {L.}~\bibnamefont {McInnes}}, \bibinfo {author} {\bibfnamefont {J.}~\bibnamefont {Healy}},\ and\ \bibinfo {author} {\bibfnamefont {J.}~\bibnamefont {Melville}},\ }\href@noop {} {\bibfield  {journal} {\bibinfo  {journal} {arXiv preprint arXiv:1802.03426}\ } (\bibinfo {year} {2018})}\BibitemShut {NoStop}%
\bibitem [{\citenamefont {Lutsko}\ and\ \citenamefont {Nicolis}(2006)}]{lutsko2006theoretical}%
  \BibitemOpen
  \bibfield  {author} {\bibinfo {author} {\bibfnamefont {J.~F.}\ \bibnamefont {Lutsko}}\ and\ \bibinfo {author} {\bibfnamefont {G.}~\bibnamefont {Nicolis}},\ }\href@noop {} {\bibfield  {journal} {\bibinfo  {journal} {Physical Review Letters}\ }\textbf {\bibinfo {volume} {96}},\ \bibinfo {pages} {046102} (\bibinfo {year} {2006})}\BibitemShut {NoStop}%
\bibitem [{\citenamefont {Beaucage}\ and\ \citenamefont {Mousseau}(2005)}]{beaucage2005nucleation}%
  \BibitemOpen
  \bibfield  {author} {\bibinfo {author} {\bibfnamefont {P.}~\bibnamefont {Beaucage}}\ and\ \bibinfo {author} {\bibfnamefont {N.}~\bibnamefont {Mousseau}},\ }\href@noop {} {\bibfield  {journal} {\bibinfo  {journal} {Physical Review B}\ }\textbf {\bibinfo {volume} {71}},\ \bibinfo {pages} {094102} (\bibinfo {year} {2005})}\BibitemShut {NoStop}%
\bibitem [{\citenamefont {Stillinger}\ and\ \citenamefont {Weber}(1985)}]{stillinger1985computer}%
  \BibitemOpen
  \bibfield  {author} {\bibinfo {author} {\bibfnamefont {F.~H.}\ \bibnamefont {Stillinger}}\ and\ \bibinfo {author} {\bibfnamefont {T.~A.}\ \bibnamefont {Weber}},\ }\href@noop {} {\bibfield  {journal} {\bibinfo  {journal} {Physical Review B}\ }\textbf {\bibinfo {volume} {31}},\ \bibinfo {pages} {5262} (\bibinfo {year} {1985})}\BibitemShut {NoStop}%
\bibitem [{\citenamefont {Luo}\ \emph {et~al.}(2024)\citenamefont {Luo}, \citenamefont {Yang}, \citenamefont {Xie}, \citenamefont {Srinivasan}, \citenamefont {Tian}, \citenamefont {Sankaranarayanan}, \citenamefont {Arslan}, \citenamefont {Yang}, \citenamefont {Mao},\ and\ \citenamefont {Wen}}]{luo2024atomistic}%
  \BibitemOpen
  \bibfield  {author} {\bibinfo {author} {\bibfnamefont {D.}~\bibnamefont {Luo}}, \bibinfo {author} {\bibfnamefont {L.}~\bibnamefont {Yang}}, \bibinfo {author} {\bibfnamefont {H.}~\bibnamefont {Xie}}, \bibinfo {author} {\bibfnamefont {S.}~\bibnamefont {Srinivasan}}, \bibinfo {author} {\bibfnamefont {J.}~\bibnamefont {Tian}}, \bibinfo {author} {\bibfnamefont {S.}~\bibnamefont {Sankaranarayanan}}, \bibinfo {author} {\bibfnamefont {I.}~\bibnamefont {Arslan}}, \bibinfo {author} {\bibfnamefont {W.}~\bibnamefont {Yang}}, \bibinfo {author} {\bibfnamefont {H.-k.}\ \bibnamefont {Mao}},\ and\ \bibinfo {author} {\bibfnamefont {J.}~\bibnamefont {Wen}},\ }\href@noop {} {\bibfield  {journal} {\bibinfo  {journal} {Carbon}\ }\textbf {\bibinfo {volume} {229}},\ \bibinfo {pages} {119538} (\bibinfo {year} {2024})}\BibitemShut {NoStop}%
\bibitem [{\citenamefont {Tersoff}(1988)}]{tersoff1988empirical}%
  \BibitemOpen
  \bibfield  {author} {\bibinfo {author} {\bibfnamefont {J.}~\bibnamefont {Tersoff}},\ }\href@noop {} {\bibfield  {journal} {\bibinfo  {journal} {Physical Review Letters}\ }\textbf {\bibinfo {volume} {61}},\ \bibinfo {pages} {2879} (\bibinfo {year} {1988})}\BibitemShut {NoStop}%
\bibitem [{\citenamefont {Marchant}\ \emph {et~al.}(2023)\citenamefont {Marchant}, \citenamefont {Caro}, \citenamefont {Karasulu},\ and\ \citenamefont {P{\'a}rtay}}]{marchant2023exploring}%
  \BibitemOpen
  \bibfield  {author} {\bibinfo {author} {\bibfnamefont {G.~A.}\ \bibnamefont {Marchant}}, \bibinfo {author} {\bibfnamefont {M.~A.}\ \bibnamefont {Caro}}, \bibinfo {author} {\bibfnamefont {B.}~\bibnamefont {Karasulu}},\ and\ \bibinfo {author} {\bibfnamefont {L.~B.}\ \bibnamefont {P{\'a}rtay}},\ }\href@noop {} {\bibfield  {journal} {\bibinfo  {journal} {npj Computational Materials}\ }\textbf {\bibinfo {volume} {9}},\ \bibinfo {pages} {131} (\bibinfo {year} {2023})}\BibitemShut {NoStop}%
\bibitem [{\citenamefont {Zhang}\ and\ \citenamefont {Piccini}(2026)}]{zhang2026exploring}%
  \BibitemOpen
  \bibfield  {author} {\bibinfo {author} {\bibfnamefont {Z.}~\bibnamefont {Zhang}}\ and\ \bibinfo {author} {\bibfnamefont {G.}~\bibnamefont {Piccini}},\ }\href@noop {} {\bibfield  {journal} {\bibinfo  {journal} {Nature Communications}\ } (\bibinfo {year} {2026})}\BibitemShut {NoStop}%
\bibitem [{\citenamefont {No{\'e}}\ \emph {et~al.}(2019)\citenamefont {No{\'e}}, \citenamefont {Olsson}, \citenamefont {K{\"o}hler},\ and\ \citenamefont {Wu}}]{noe2019boltzmann}%
  \BibitemOpen
  \bibfield  {author} {\bibinfo {author} {\bibfnamefont {F.}~\bibnamefont {No{\'e}}}, \bibinfo {author} {\bibfnamefont {S.}~\bibnamefont {Olsson}}, \bibinfo {author} {\bibfnamefont {J.}~\bibnamefont {K{\"o}hler}},\ and\ \bibinfo {author} {\bibfnamefont {H.}~\bibnamefont {Wu}},\ }\href@noop {} {\bibfield  {journal} {\bibinfo  {journal} {Science}\ }\textbf {\bibinfo {volume} {365}},\ \bibinfo {pages} {eaaw1147} (\bibinfo {year} {2019})}\BibitemShut {NoStop}%
\end{thebibliography}

\clearpage

\onecolumngrid



\section*{Supplementary material}

# Supplemental Material for: Information entropy is a general-purpose collective variable for enhanced sampling

**Xiangrui Li,**[1] **Daniel Schwalbe-Koda**[1,*]

[1] Department of Materials Science and Engineering, University of California, Los Angeles, CA, USA
[*] E-mail: dskoda@ucla.edu

## S1. Supplementary Methods

### S1.1. Choice of simulation parameters for $\delta\mathcal{H}$-MetaD

Our method uses the conventional MetaD parameters — bias height, bias width ($\sigma$), and bias deposition frequency (stride $N$) — along with a new parameter introduced here, the kernel bandwidth ($h$). Among these parameters, $\sigma$ and $h$ are the most important because they strongly control exploration behavior and are coupled to each other. The bias width $\sigma$ sets the spatial extent of the repulsive hills along the CV, and therefore controls how broadly the free-energy surface is pushed. The kernel bandwidth $h$ controls the overlap between $\delta\mathcal{H}$ distributions: a larger $h$ yields greater overlap, so two configurations are more likely to be treated as similar. As a result, smaller $h$ generally requires larger $\sigma$, and vice versa. When $h$ is small, configurations appear more dissimilar in $\delta\mathcal{H}$ space, so a wider bias is needed for effective bias transfer. In practice, $\sigma$ is obtained by analyzing the characteristic $\delta\mathcal{H}$ distance $D$ between the initial state and final metastable states, which is often unknown *a priori*. By contrast, $h$ can be estimated more systematically when representative configurations are available. This coupling can be summarized as

$$\sigma \propto \frac{D}{h} \tag{S1}$$

### S1.2. Simulation details

**Alanine dipeptide (Ala2) and alanine tetrapeptide (Ala4).** We performed 10 ns unbiased and enhanced simulations for both molecules in vacuum at 300 K with a Langevin thermostat, using *openff-2.0.0.offxml* from OpenFF[1] and OpenMM[2]. The Gaussian bias height was 1.2 kJ/mol for both systems. The Gaussian width ($\sigma$) is reported in extensive units of nats, which can be converted to the PySAGES[3] native unit nats by multiplying by the system size. For Ala2 the Gaussian width was set to 0.045 nats; for Ala4, it was set to 0.38 nats to enable broader exploration of multiple metastable states. Because the two molecules are structurally similar, the entropy-change bandwidth was set to 1 Å$^{-1}$ in both cases.

**Copper.** We performed 1 ns-long simulations in the NPT ensemble at 1100 K and 0 bar using LAMMPS (29 Aug 2024)[4]. The system contains 500 atoms described by the Mishin EAM potential[5]. To generate a supercooled liquid reference, we melted the system at 2000 K and quenched it back to 1100 K at 1 K/ps, then it was equilibrated under NPT ensembles for 100 ps at 1100 K and 0 bar before enhanced sampling. The bandwidth was set to 0.025 Å$^{-1}$ by directly comparing the $\delta\mathcal{H}$ distributions of FCC and amorphous structures at the same temperature. Bias heights between 0.05 and 0.1 eV, with 0.008 nats bias width were successful in driving the transformations. These values are also

interpretable considering the typical energy barriers[6,7] and ranges of $\delta\mathcal{H}$ transformation between -3 and 0 nats.

**Silicon.** We performed 5 ns-long simulations of 512 atoms using the SW potential.[8] Each system was first relaxed in the NPT ensemble at the target temperature and 0 bar to preserve crystallinity, and then switched to the NVT ensemble to maintain crystalline density. We generated a supercooled reference by melting at 2500 K and quenching to the target temperature at 1 K/ps. The bandwidth was set to 0.035 Å$^{-1}$ from direct comparison between crystalline and amorphous silicon $\delta\mathcal{H}$ distributions at the same temperature. We tested target temperatures of 1200 K and 1250 K, bias height of 0.1 eV, and bias widths between 0.0078125 and 0.03125 nats, where trajectory with bias width equal to 0.0078125 nats successfully nucleate. These values were chosen based on the nucleation energy barriers[9] and $\delta\mathcal{H}$ distributions under different states. At 1200 K, unbiased simulation already exhibits a nucleation pathway, and biased simulation either reproduces nucleation or exploits a glass transition pathway. At higher temperature (1250 K), the supercooled liquid remains stable under 10 ns unbiased simulation, while biased simulation can only drive the system into a glass state. In both cases, $\delta\mathcal{H}$-MetaD is capable of exploring rare events and showcasing the $\delta\mathcal{H}$ distribution shift. Adjusting the bias width allows us to explore configurational space with different scope, where larger bias width always leads to a glass state, implying that the glass state $\delta\mathcal{H}$ distribution is further from the reference state compared to the crystalline state.

**Graphite.** We performed 1 ns simulations in the NPT ensemble using the Tersoff potential,[10] at 3000 K and 100 GPa. The 512-atom system was initialized in the graphite phase. The bandwidth was set to 0.25 Å$^{-1}$ by directly comparing graphite and diamond $\delta\mathcal{H}$ distributions. Because the Tersoff potential requires more extreme conditions to induce nucleation, we used bias height of 0.1 eV and bias width of 0.0125 nats which were chosen based on the reported activation energy[11] and $\delta\mathcal{H}$ distributions between two states.

### S1.3. Intermediate states analysis

The formation of intermediate states in both the copper and silicon examples demonstrate that there is a structural dissimilarity to the liquid and crystalline states, whereas polyhedral template matching[12] identifies them as amorphous. We then applied $\delta\mathcal{H} = -1.5$ as a threshold to classify high $\delta\mathcal{H}$ atoms and low $\delta\mathcal{H}$ atoms from unsupervised $\delta\mathcal{H}$ distribution and analyzed their local density fluctuations along multiple trajectories. We found that intermediate states form due to the local density fluctuation (Figs. S4a-c, S5a,b), where ordered (low $\delta\mathcal{H}$) atoms densify. Interestingly, this ordered-atom densification diverges into two different routes in silicon example: one (orange line) exhibits a metastable intermediate before nucleating, while the other (green line) passes through a transient intermediate state and transforms into glass. We also provide snapshots at different states for an intuitive understanding of the $\delta\mathcal{H}$ distribution (Figs. S4d-g, S5c-f), that crystalline have aggregated low $\delta\mathcal{H}$ spatial distribution while glass is dispersed. These results demonstrate that $\delta\mathcal{H}$ serves not only as a CV, but also as a sensitive descriptor capable of capturing subtle structural changes in the nucleation process.

### S1.4. Details of 2-step $\delta\mathcal{H}$-MetaD sampling

When geometrical coordinates are remapped into $\delta\mathcal{H}$ space, metastable configurations with larger $\delta\mathcal{H}$ often correspond to extended and reshaped valleys in $E(\delta\mathcal{H})$ (see Fig. 1c). Traversing these valleys can require substantially longer simulations. A practical strategy is to expand the reference

dataset $\{\mathbf{X}\}$ with an intermediate set $\mathbf{X}_{\text{inter}}$. Following this idea, we test a two-step MetaD protocol for graphite at 1500 K while keep other parameters same to the graphite example. At this temperature, direct graphite-to-diamond transitions remain difficult because of the high barrier; additionally, the lower temperature increases the $\delta\mathcal{H}$ distance between graphite and diamond, which further slows convergence. As shown in Fig. S7 (left), the first 1 ns MetaD step reaches a buckled graphite structure. In the second step, we add this buckled graphite frame to the reference dataset, which then enables sampling of a diamond nucleus forming within graphite (Fig. S7, right).

## S2. Supplementary Figures

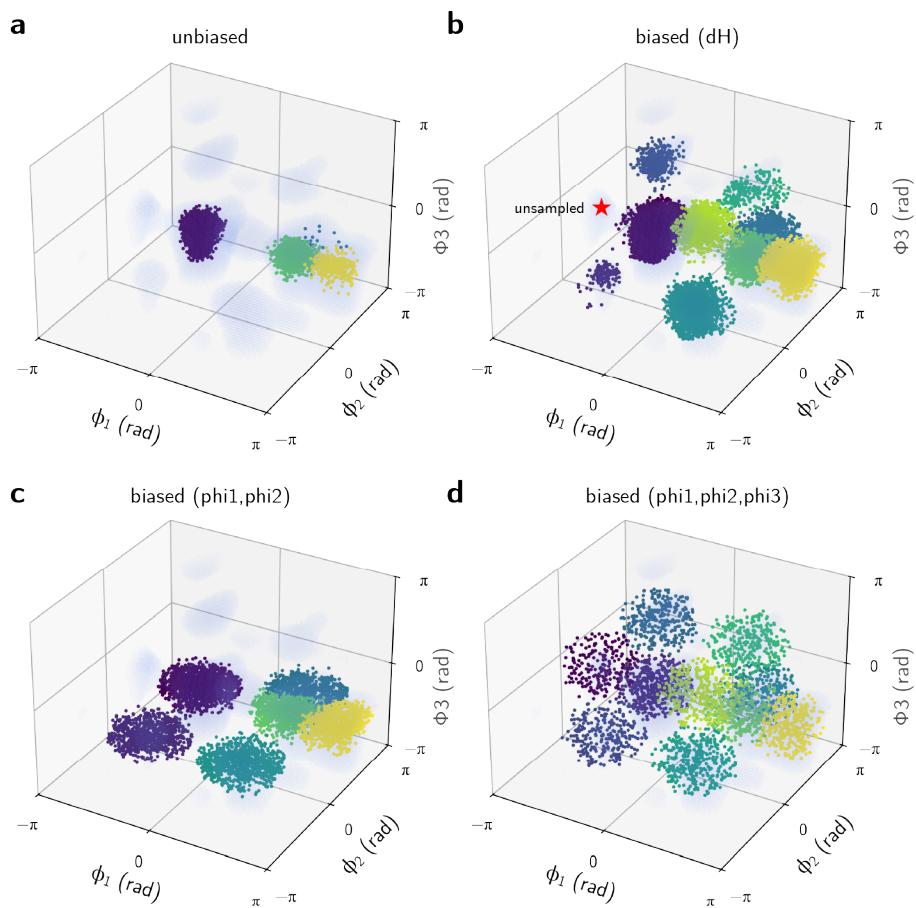

**Fig.** S1: Clustering of sampled states from unbiased, MetaD-$\delta\mathcal{H}$, MetaD-$\phi1$, $\phi2$, and MetaD-$\phi_1$, $\phi_2$, $\phi_3$ trajectories. Unbiased simulation is trapped in three states, while $\phi_1$, $\phi_2$ biased trajectory is trapped in the $\phi_1$, $\phi_2$ plane. Only $\delta\mathcal{H}$ and $\phi_1$, $\phi_2$, $\phi_3$ biased trajectories can extend the sampling of metastable state across $\phi_3$ axis, and $\delta\mathcal{H}$ biased trajectory left one unsampled metastable state.

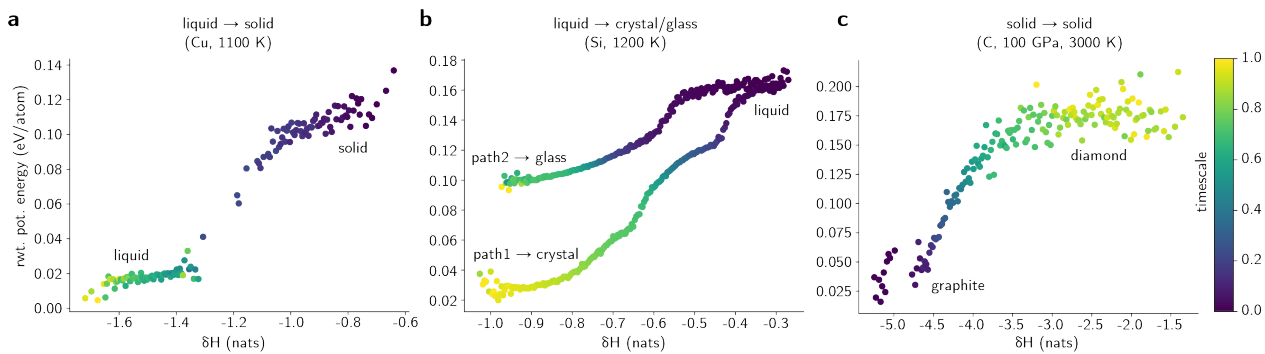

**Fig.** S2: Compact 2D plot of Boltzmann reweighted potential energy versus averaged $\delta\mathcal{H}$, where initial states and final states are labeled on the figure. For the silicon, two lines represent the results from two biased trajectory that lead to different final state. Color represents the temporal sequence of sampled states.

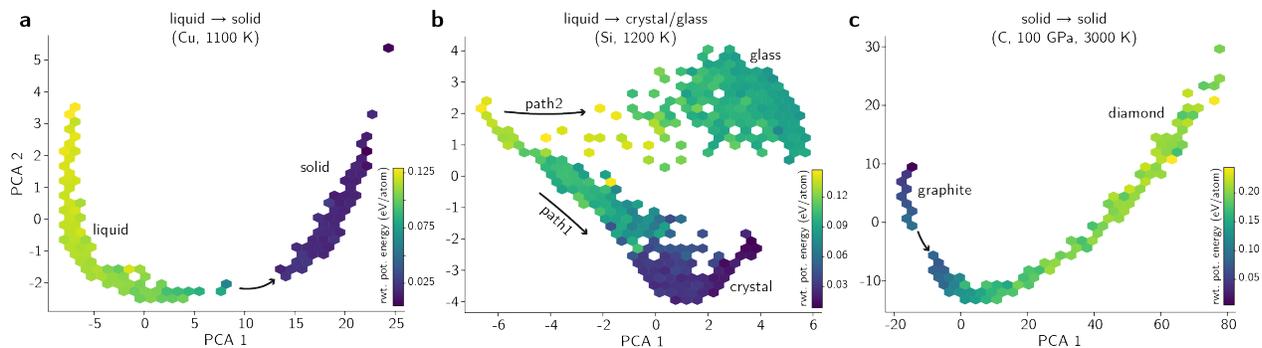

**Fig.** S3: Boltzmann-reweighted potential energy profile on PCA reduced dimension of Wasserstein-1 distance matrix of $\{\delta\mathcal{H}\}$ across multiple trajectories. Here $\{\delta\mathcal{H}\} = \{\delta\mathcal{H}(\mathbf{Y}_{ij}|\{\mathbf{X}\})\}$ is the set of $\delta\mathcal{H}$ computed over all frames $i$ and trajectories $j$.

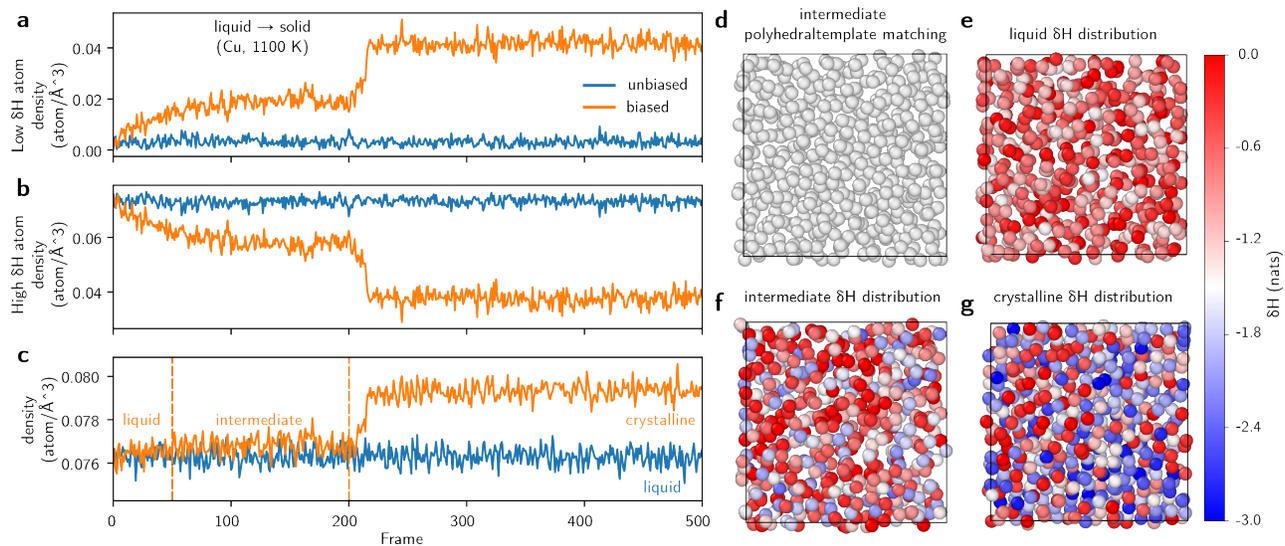

**Fig.** S4: **a-c** Low $\delta\mathcal{H}$ atom, high $\delta\mathcal{H}$ atom and global densities of copper along two trajectories. $\delta\mathcal{H} = -1.5$ is chosen as the threshold for high $\delta\mathcal{H}$ and low $\delta\mathcal{H}$ atom classification. **d** Polyhedral template matching of intermediate state snapshot. **e-g** $\delta\mathcal{H}$ color coded snapshot of liquid, intermediate and crystalline state snapshots.

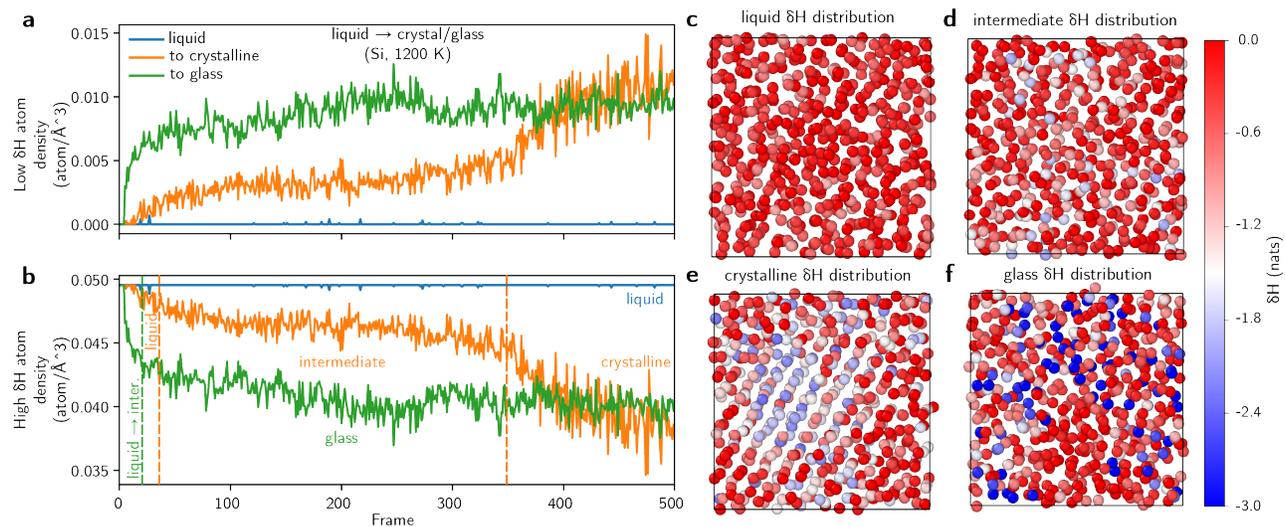

**Fig.** S5: **a-b** Low $\delta\mathcal{H}$ atom, high $\delta\mathcal{H}$ atom and global densities of silicon along one unbiased trajectory and two biased trajectories with different final states. $\delta\mathcal{H} = -1.5$ is chosen as the threshold for high $\delta\mathcal{H}$ and low $\delta\mathcal{H}$ atom classification. **c-f** $\delta\mathcal{H}$ color coded snapshot of liquid, intermediate, crystalline and glass state snapshots.

6

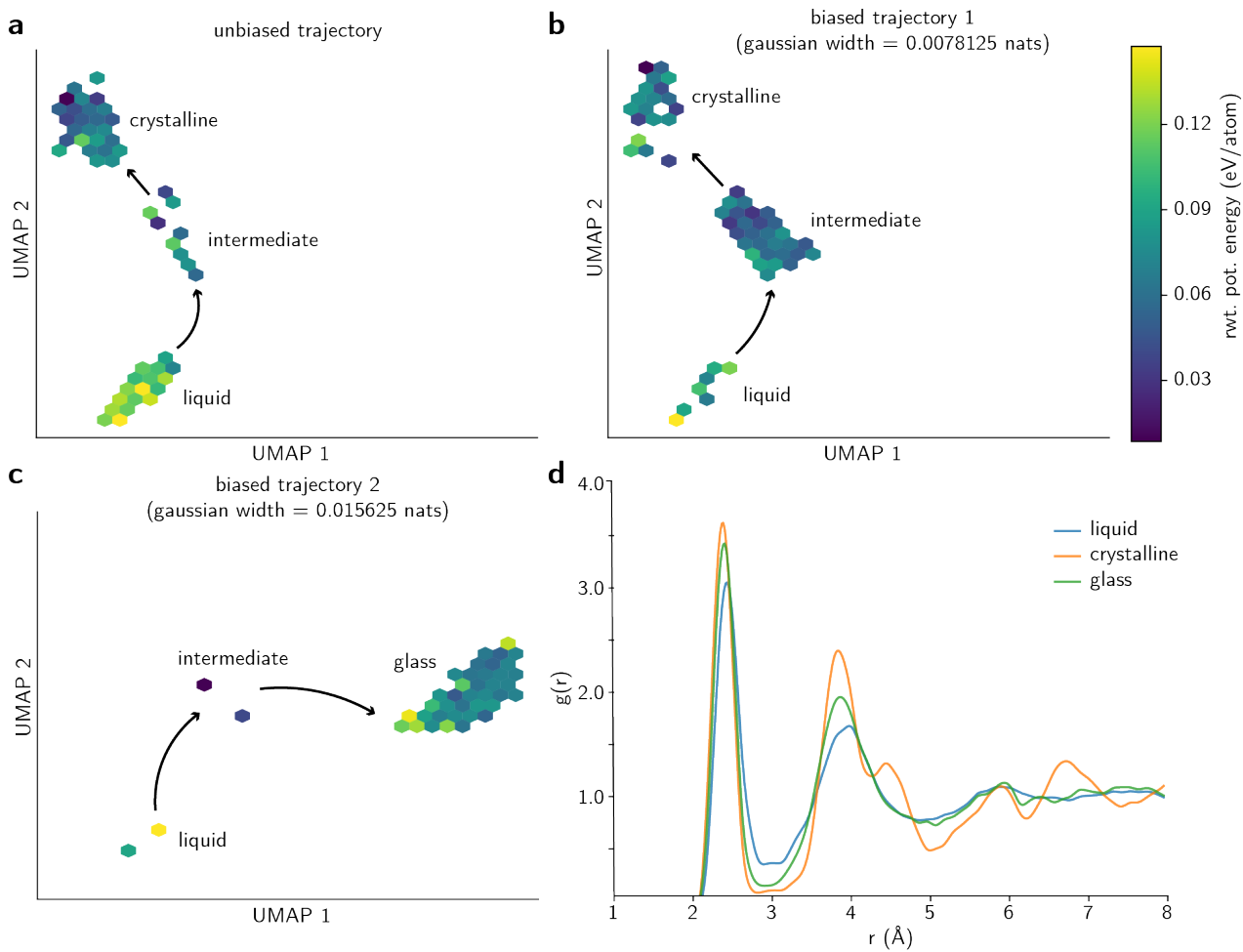

**Fig.** S6: Decomposed individual reaction pathways of the unbiased trajectory from liquid to crystalline, and two biased trajectories with different bias widths that transfer from liquid to crystalline and glass shown in Fig. 3h. Right below panel shows the radial distribution function of different states. With smaller bias width (biased trajectory 1), the $\delta\mathcal{H}$ distribution shifts to intermediate state faster than unbiased simulation, showcasing the effect of bias force, while larger bias width (biased trajectory 2) glide through intermediate state without being trapped. We assume that intermediate state is crucial to the crystallization pre-ordering and biased trajectory 2 skips this step by directly nucleating into glass.

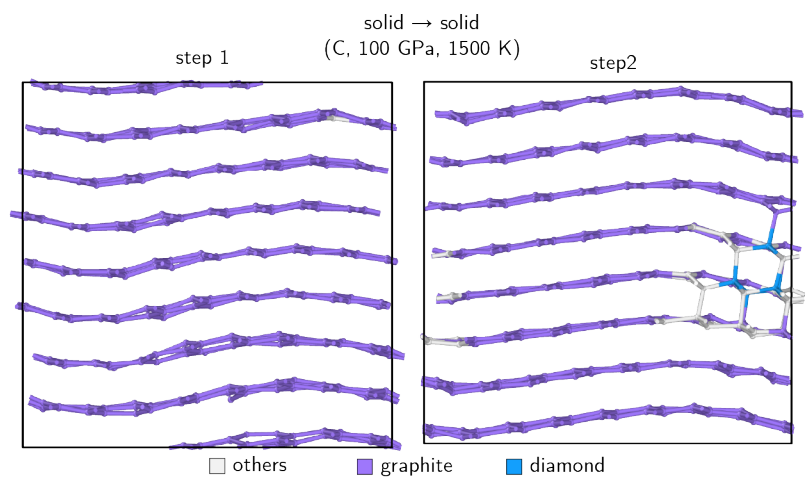

**Fig.** S7: Final frame from the first-step MetaD trajectory, showing buckled graphite (left), and final frame from the second-step trajectory, showing graphite containing a diamond nucleus (right).

# References

[1] Wagner, J., Thompson, M., Dotson, D., Rodríguez-Guerra, J. *et al.* openforcefield/openff-forcefields: Version 2.0. 0" sage". *Zenodo* (2021).

[2] Eastman, P. *et al.* Openmm 8: molecular dynamics simulation with machine learning potentials. *The Journal of Physical Chemistry B* **128**, 109–116 (2023).

[3] Zubieta Rico, P. F. *et al.* Pysages: flexible, advanced sampling methods accelerated with gpus. *npj Computational Materials* **10**, 35 (2024).

[4] Thompson, A. P. *et al.* Lammps-a flexible simulation tool for particle-based materials modeling at the atomic, meso, and continuum scales. *Computer Physics Communications* **271**, 108171 (2022).

[5] Mishin, Y., Mehl, M. J., Papaconstantopoulos, D. A., Voter, A. F. & Kress, J. D. Structural stability and lattice defects in copper: Ab initio, tight-binding, and embedded-atom calculations. *Physical Review B* **63**, 224106 (2001).

[6] Bechelli, S. *et al.* Free energy of nucleation and interplay between size and composition in cuni systems. *The Journal of Physical Chemistry B* **121**, 8558–8563 (2017).

[7] Desgranges, C. & Delhommelle, J. Effect of the composition on the free energy of crystal nucleation for cupd nanoalloys. *The Journal of Physical Chemistry C* **120**, 27657–27664 (2016).

[8] Stillinger, F. H. & Weber, T. A. Computer simulation of local order in condensed phases of silicon. *Physical Review B* **31**, 5262 (1985).

[9] Beaucage, P. & Mousseau, N. Nucleation and crystallization process of silicon using the stillinger-weber potential. *Physical Review B* **71**, 094102 (2005).

[10] Tersoff, J. Empirical interatomic potential for carbon, with applications to amorphous carbon. *Physical Review Letters* **61**, 2879 (1988).

[11] Fahy, S., Louie, S. G. & Cohen, M. L. Pseudopotential total-energy study of the transition from rhombohedral graphite to diamond. *Physical Review B* **34**, 1191 (1986).

[12] Larsen, P. M., Schmidt, S. & Schiøtz, J. Robust structural identification via polyhedral template matching. *Modelling and Simulation in Materials Science and Engineering* **24**, 055007 (2016).


\AtBeginShipoutNext{\AtBeginShipoutDiscard}

\end{document}